\documentclass[onecolumn,floatfix,aps,nofootinbib]{revtex4}
\pdfoutput=1
\usepackage[dvips]{epsfig}
\usepackage[english]{babel}
\usepackage[utf8]{inputenc}
\usepackage[T1]{fontenc}
\usepackage{lipsum}       
\usepackage{float}
\usepackage{bbm}
\usepackage{verbatim}
\usepackage{array}
\usepackage{bm} 
\usepackage{amsmath}
\usepackage{yfonts}
\usepackage{amsthm}
\usepackage{amsmath,amscd}
\usepackage{pst-plot}
\usepackage{slashed} 
\usepackage{tikz-cd}
\usepackage{accents}
\usepackage{amsfonts}
\usepackage{graphicx}
\usepackage{amssymb}

\usepackage{cases}

\usepackage{array}
\newcolumntype{C}[1]{>{\centering\arraybackslash}m{#1}}

\usepackage{indentfirst} 
\usepackage[titletoc]{appendix}
\usepackage{appendix}
\usepackage{subeqnarray}
\usepackage{indentfirst} 

\usepackage{xcolor}
\usepackage{calrsfs}

\usepackage{wasysym}
\usepackage[all]{xy}
\usepackage{tikz-cd}

\usepackage{amsmath,amsfonts,amssymb,amsthm,mathrsfs,bbm,braket}

\numberwithin{equation}{section}

\usepackage{bbold}

\usepackage{extarrows}
\usepackage{setspace}
\begin{document}

\title{$\mathrm{ISIM}(2)$ gravitational gauge model via Soft Algebra formalism}  
    
    \author{J. M. B. Matzenbacher}
   \email{j.matzenbacher@unesp.br}
   \affiliation{S\~ao Paulo State University (UNESP), School of Engineering and Sciences, Guaratinguet\'a - Brazil.}

    \author{J. M. Hoff da Silva}
   \email{julio.hoff@unesp.br}
   \affiliation{S\~ao Paulo State University (UNESP), School of Engineering and Sciences, Guaratinguet\'a - Brazil.}

\begin{abstract}
    This work presents a comprehensive account of a gravitational model based on the gauging of the $\mathrm{ISIM}(2)$ Lorentz subgroup parameters through the so-called soft algebra formalism. After presenting a closed form for $\mathfrak{sim}(2)$ and its inhomogeneous extension, we proceed to develop a complete torsionless gravitational model, culminating in an invariant action.
\end{abstract}		

    \maketitle
   
    \section{Introduction}
    Gauge theories constitute the foundational framework for describing the fundamental interactions of nature, providing a unified language that spans from classical field theories to the Standard Model of particle physics. In this formulation, continuous local symmetries are leveraged to derive gauge fields and the conservation laws that govern physical dynamics. Beyond the internal symmetry groups of particle physics, this gauge paradigm has been successfully extended to the gravitational domain. By gauging the complete Poincaré group, as pioneered by T. W. B. Kibble \cite{Kibble}, it is shown that both the non-holonomic vielbein contributions and the spin connection emerge naturally as compensating fields associated with local Lorentz symmetry combined with general coordinate transformations of spacetime. As shown in \cite{BlagojevicA, BlagojevicB}, this procedure leads to Riemann-Cartan geometry, characterized by the presence of torsion, thereby establishing a robust generalization of general relativity and fully integrating gravity into the framework of classical field theories.

    In subsequent decades, the program of interpreting gravity from a gauge-theoretic perspective was significantly deepened, particularly through developments in supersymmetry and supergravity. Within this framework, the geometric formalism of group manifolds, often associated with the notion of soft gauge algebras, which are characterized by structure functions instead of constants \cite{FV-P}, established itself as a powerful tool. This approach provides a mathematically sound formulation that is compatible, in differential-geometric terms, with both base manifold diffeomorphisms and the internal symmetries of gauge groups \cite{geoperspec, Castellani1, Neeman}. When applied to the Poincaré group, this framework coincides with the geometric structure of the Einstein-Cartan gravity model, identifying the spin connection and tetrad fields as independent dynamical variables from which geometric entities, such as curvature and torsion, are constructed as non-trivial fields.
    
    Spacetime symmetries remain central to fundamental physics, dictating the geometric properties of spacetime and field dynamics. While the Poincaré group ensures the consistency of the Standard Model, its empirical shortcomings (such as the origin of neutrino masses, baryon asymmetry, and the nature of the dark components of the universe) motivate the exploration of conceptual extensions where exact Lorentz invariance is reduced to an effective symmetry. In this context, Cohen and Glashow proposed Very Special Relativity ($\mathrm{VSR}$) \cite{VSR}, in which the underlying spacetime symmetry is restricted to proper subgroups of $\mathrm{SO}(1,3)$, emphasizing the roles played by the subgroups $\mathrm{HOM}(2)$ and $\mathrm{SIM}(2)$. Without requiring the full Lorentz group, these two subgroups are capable of reproducing all key kinematic effects of special relativity, such as the negative result of the Michelson-Morley experiment, time dilation, relativistic velocity addition, and the maximality of the speed of light in vacuum \cite{GABRIEL, Salinas}. The $\mathrm{SIM}(2)$ subgroup is of particular interest relative to $\mathrm{HOM}(2)$, due to its additional spatial rotation generator associated with a residual $\mathrm{SO}(2)$ symmetry. Furthermore, interest in the $\mathrm{SIM}(2)$ group is also bolstered by an intrinsic feature of $\mathrm{VSR}$: Lorentz invariance violation does not stem from the introduction of external background fields or fine-tuned breaking parameters, as typically occurs in standard symmetry-breaking frameworks, but rather emanates from the direct replacement of the fundamental symmetry group with one of its proper subgroups. In particular, $\mathrm{VSR}$ invariance under the $\mathrm{SIM}(2)$ group implies the existence of a privileged spacetime direction, characterized by the invariance (up to a scaling factor) of a specific class of null vectors. In this spirit, a model for an anisotropic special relativity based on $\mathrm{SIM}(2)$ symmetry was explored in \cite{Anisotropia2}. While preserving the validity of the classical tests of special relativity, this framework enables a range of unconventional geometric structures through the controlled introduction of anisotropies.
    
    Moving from a flat spacetime background to a fully dynamical gravitational scenario requires localizing these novel symmetry structures. To this end, the main objective of this paper is to formulate a consistent gravitational model for the inhomogeneous $\mathrm{SIM}(2)$ group, denoted as $\mathrm{ISIM}(2)$ and formally defined as the semi-direct product of $\mathrm{SIM}(2)$ with the Abelian group of spacetime translations. We construct our gravitational model within the mathematical framework of soft gauge algebras \cite{soft}, introduced by M. F. Sohnius. This framework allows the standard inhomogeneous Lie algebra structure constants to be replaced by spacetime-dependent structure functions, thereby accommodating the intricate interplay between base manifold diffeomorphisms and the internal symmetries of the model. A salient feature of Sohnius's formalism is its capacity to recover non-relativistic Newton-Cartan gravity when formulated as a gauge theory of the Bargmann algebra, which is defined as the central extension of the Galilei group \cite{Andringa et al}.
    
    Although prior works have associated a gravitational version of $\mathrm{VSR}$ \cite{GibbonsisFinsler, KouretsisinFinsler}, they approached the discussion from the standpoint of a Finslerian metric geometry. The present paper aims to provide a comprehensive, bottom-up derivation of the emergent geometric facets of a generalized very special relativity theory by gauging the $\mathrm{VSR}$ symmetries. Our manuscript is organized around two primary developments that advance this program. First, in Section \ref{Sec:2}, we revisit the derivation of the $\mathrm{ISIM}(2)$ algebra as a contraction of the Poincaré algebra, a result previously established by the authors in \cite{primos}. The primary objective of this section is to explore a three-dimensional representation of the $\mathrm{ISIM}(2)$ generator algebra. This investigation enables us to define and interpret the $\mathrm{VSR}$ boosts and rotations upon which the gauge model is constructed. Next, in Section \ref{Sec:3}, we present the construction of the gravitational model, which culminates in the proposal of a compatible functional action. Detailed calculations of several relevant relations are relegated to the appendices.

    \section{From Poincaré to $\mathrm{ISIM}(2)$: a motivation for a three dimensional representation}\label{Sec:2}
    
    In this section, we construct a three-dimensional representation of the $\mathfrak{isim}(2)$ Lie algebra. Our strategy is to revisit the realization of the $\mathrm{SIM}(2)$ group as a contraction of the Lorentz group, adopting a perspective that complements the approach previously explored in \cite{primos}. In particular, the proposed construction will prove not only to elucidate a possible physical interpretation of the $\mathrm{SIM}(2)$ group, but also to be advantageous for the subsequent gauging of the resulting algebra (see Section \ref{Sec:3}).
    
    \subsection{A concise review of In\"on\"u-Wigner contraction procedure}\label{subsec: 2.A}
    
    Let $G$ be a Lie group. Every element $g \in G$ corresponds to a point on the underlying manifold, and the set of all tangent vectors—defined via equivalence classes of smooth curves passing through $g$—constitutes the tangent space to $G$ at this point, denoted here as $T_gG$. Beyond its vector space structure, the specific tangent space at the identity element $\mathbb{1}_G\in G$ is endowed with an additional algebraic structure induced by the Lie bracket of vector fields. For instance, given an arbitrary basis $\{I_i\}$ ($i=1,2,\dots,n$) for $T_{\mathbb{1}_G}G$, this structure is encoded in the commutation relations\footnote{Throughout this work, we adopt Einstein's summation convention.}
    \begin{equation}
        \label{algebra}
        [I_i,I_j] = C_{i\phantom{k}j}^{\phantom{i}k}I_k,
    \end{equation}
    where $C_{i\phantom{k}j}^{\phantom{i}k}$ are the so-called structure constants of the Lie algebra associated with $G$ \cite{Isham,Nakahara}. The choice of a basis for the tangent space is not unique, but rather defined up to a non-singular linear transformation. Let us consider a matrix $U \in GL(n,\mathbb{R})$ such that a new set of basis elements is obtained as follows:
    \begin{equation}
        \label{itiltoi}
        \tilde{I}_i = U_i^{\phantom{i}j}I_j.
    \end{equation}
    Since the Lie algebra must be basis-independent, the introduction of (\ref{itiltoi}) into (\ref{algebra}) allows us to recast the latter by means of the new basis:
    \begin{equation}
        \label{newstrucconst}
        [\tilde{I}_i,\tilde{I}_j] = \tilde{C}_{i\phantom{k}j}^{\phantom{i}k}\tilde{I}_k,
    \end{equation}
    where $\tilde{C}_{i\phantom{k}j}^{\phantom{i}k}$ must be written in terms of the old structure constant as
    \begin{equation}
        \tilde{C}_{i\phantom{k}j}^{\phantom{i}k} = U_j^{\phantom{i}l}U_i^{\phantom{j}m} C_{l\phantom{m}m}^{\phantom{l}n}(U^{-1})_n^{\phantom{i}k}.
    \end{equation}
   Furthermore, any group element $g \in G$ can be locally expressed as $g = \exp\{i\theta^iI_i\}$, where $\{\theta^i\}$ ($i=1,2,\dots,n$) corresponds to the group parameters. Since the representation of $g \in G$ must be basis-independent, when recast in terms of $\{\tilde{I}_i\}$, we have $g = \exp\{i\tilde{\theta}^i\tilde{I}_i\}$. This leads to the identification $\tilde{\theta}^i = \theta^j(U^{-1})_j^{\phantom{i}i}$, which relates the new group parameters to the original ones.
   
    Provided that $U$ is non-singular, the proposed transformation does not modify the structure of the group $G$. The crucial aspect of the In\"on\"u-Wigner contraction procedure lies precisely in abandoning this restriction. By allowing the transformation matrix to become singular in a controlled limiting process, one can construct a new Lie group (and hence a new Lie algebra) that is no longer isomorphic to the original one. The contraction scheme is entirely governed by the first In\"on\"u-Wigner theorem, which states that every Lie group can be contracted with respect to any of its continuous subgroups, and only with respect to such subgroups \cite{Contracmethod, Feza}. The key idea is to express the non-singular matrix $U$ by introducing a real-valued parameter $\varepsilon$, as follows:
    \begin{equation}
        \label{normal-form1}
        U_i^{\phantom{i}j} = u_i^{\phantom{i}j} + \varepsilon w_i^{\phantom{i}j}, 
    \end{equation}
    with the proviso that $\det{U} \neq 0$ for $\varepsilon\neq 0$, while $\det{U} = 0$ in the limit $\varepsilon \to 0$. Henceforth, we shall assume that both matrices $u_i^{\phantom{i}j}$ and $w_i^{\phantom{i}j}$ are written in the so-called `normal form', characterized by the block decomposition
    \begin{equation}
        \label{normal-form}
        u = \begin{pmatrix}
            \mathbb{1}_{r\times r} & 0_{r\times (n-r)}\\
            0_{(n-r)\times r} & 0_{(n-r)\times (n-r)}
        \end{pmatrix}\qquad \text{and} \qquad w = \begin{pmatrix}
            v_{r\times r} & 0_{r\times(n-r)}\\
            0_{(n-r)\times r} & \mathbb{1}_{(n-r)\times(n-r)} 
        \end{pmatrix},
    \end{equation}
    where $\mathbb{1}$ and $0$ denote the identity and zero matrices of the indicated dimensions, respectively, $v$ is an arbitrary $r\times r$ matrix, and the integer $r$ satisfies the constraint $0<r<n$. For completeness, it is worth noting that the normal-form structure displayed in (\ref{normal-form}) is not always attainable (see \cite{Convergencetosingular}). However, for the purposes of the present work, we shall show that this construction is indeed suitable for the case under consideration.

    The employment of (\ref{normal-form1}) into relation (\ref{itiltoi}) yields 
    \begin{eqnarray}
        \label{eitaeita}
        \tilde{I}_{1,\mu} &=& I_{1,\mu} + \varepsilon v_\mu^{\phantom{\mu}\nu}I_{1,\nu}, \qquad (\mu,\nu = 1,2,...,r),\\
        \label{eitaeita2}
        \tilde{I}_{2,\lambda} &=& \varepsilon I_{2,\lambda}, \qquad (\lambda = r+1, r+2, ..., n ).
    \end{eqnarray}
    The subscripts `1' and `2' are used to distinguish the two sectors associated with the matrices $u$ and $w$ upon which the generators $\{I_i\}$ act. Specifically, the label `1' refers to the first $r$ generators, while `2' corresponds to the remaining $n-r$ generators. Greek letters are then employed to label all generators belonging to each sector. Given that $\det{U} \neq 0$, expressions (\ref{eitaeita}) and (\ref{eitaeita2}) admit an equivalent inverse relation, written as
    \begin{eqnarray}
        \label{eitaeitainv}
        I_{1,\mu} &=& \tilde{I}_{1,\mu} - \varepsilon v_\mu^{\phantom{\mu}\nu}\tilde{I}_{1,\nu}, \qquad (\mu,\nu = 1,2,...,r);\\
        \label{eitaeita2inv}
        I_{2,\lambda} &=& \frac{1}{\varepsilon} \tilde{I}_{2,\lambda}, \qquad (\lambda = r+1, r+2, ..., n ).
    \end{eqnarray}

   The contraction procedure is completed by analyzing the new set of group structure constants defined in (\ref{newstrucconst}) in the limiting regime $\varepsilon \to 0$. We first analyze the commutation relation $[\tilde{I}_{1,\mu},\tilde{I}_{1,\nu}]$; the extension to the remaining Lie bracket combinations, namely $[\tilde{I}_{1,\mu},\tilde{I}_{2,\lambda}]$ and $[\tilde{I}_{2,\lambda},\tilde{I}_{2,\rho}]$, then follows straightforwardly. Using (\ref{eitaeita}) and (\ref{eitaeita2}), the commutator $[\tilde{I}_{1,\mu},\tilde{I}_{1,\nu}]$ can be expanded as
    \begin{equation}
        [\tilde{I}_{1,\mu},\tilde{I}_{1,\nu}] = [I_{1,\mu},I_{1,\nu}] + \varepsilon v_\nu^{\phantom{\nu}\alpha}[\tilde{I}_{1,\mu},\tilde{I}_{1,\alpha}] + \varepsilon v_\mu^{\phantom{\mu}\alpha}[\tilde{I}_{1,\alpha},\tilde{I}_{1,\nu}] + \mathcal{O}(\varepsilon^2). 
    \end{equation}
    With the aid of expressions (\ref{eitaeitainv}) and (\ref{eitaeita2inv}), together with the Lie algebra relation (\ref{algebra}), one finds
    \begin{equation}
        \label{ItilcomItilfinal}
        [\tilde{I}_{1,\mu},\tilde{I}_{1,\nu}] = C_{1,\mu\phantom{1,\alpha}1,\nu}^{\phantom{1,\mu}1,\alpha} \tilde{I}_{1,\alpha} + \frac{1}{\varepsilon}C_{1,\mu\phantom{2,\alpha}1,\nu}^{\phantom{1,\mu}2,\alpha} \tilde{I}_{2,\alpha} + v_\nu^{\phantom{\nu}\beta}C_{1,\mu\phantom{2,\alpha}1,\beta}^{\phantom{1,\mu}2,\alpha} \tilde{I}_{2,\alpha} + v_\mu^{\phantom{\nu}\beta}C_{1,\beta\phantom{2,\alpha}1,\nu}^{\phantom{1,\beta}2,\alpha} \tilde{I}_{2,\alpha} + \mathcal{O}(\varepsilon),
    \end{equation}
    which tells us that for this expression to be convergent under the proposed limit, it is necessary that the constraint 
    \begin{equation}
        \label{conditionimportante}
        C_{1,\mu\phantom{2,\alpha}1,\nu}^{\phantom{1,\mu}2,\alpha} = 0
    \end{equation}
    be satisfied from the outset. Recall that the new basis of generators, when connected with $\{I_i\}$ in the singular limit, must span a new Lie group structure. Therefore, it is expected that the set $\{\tilde{I}_i\}$ serves as a basis for the new Lie algebra
    \begin{equation}
        \label{newnolimit}
        [\tilde{I}_{1,\mu},\tilde{I}_{1,\nu}]_{\varepsilon \to 0} = c_{1,\mu\phantom{1,\lambda}1,\nu}^{\phantom{1,\mu}1,\lambda}\tilde{I}_{1,\lambda} + c_{1,\mu\phantom{2,\lambda}1,\nu}^{\phantom{1,\mu}2,\lambda}\tilde{I}_{2,\lambda}.
    \end{equation}
    Assuming $\varepsilon\to 0$, and comparing (\ref{ItilcomItilfinal}) with (\ref{newnolimit}), one finds
    \begin{equation}
        \label{resultcontra1}
        c_{1,\mu\phantom{1,\lambda}1,\nu}^{\phantom{1,\mu}1,\lambda} = C_{1,\mu\phantom{1,\lambda}1,\nu}^{\phantom{1,\mu}1,\lambda}\qquad \text{and} \qquad c_{1,\mu\phantom{2,\lambda}1,\nu}^{\phantom{1,\mu}2,\lambda} = 0 .
    \end{equation}
    Repeating the same analysis for $[\tilde{I}_{1,\mu},\tilde{I}_{2,\lambda}]$ and $[\tilde{I}_{2,\lambda},\tilde{I}_{2,\rho}]$, one obtains, respectively,
    \begin{eqnarray}
        [\tilde{I}_{1,\mu},\tilde{I}_{2,\lambda}] = C_{1,\mu\phantom{2,\lambda}1,\nu}^{\phantom{1,\mu}2,\lambda}\tilde{I}_{2,\lambda} + \varepsilon C_{1,\mu\phantom{1,\lambda}1,\nu}^{\phantom{1,\mu}1,\lambda}\tilde{I}_{1,\lambda} + \mathcal{O}(\varepsilon),
    \end{eqnarray}
    and
    \begin{eqnarray}
        [\tilde{I}_{2,\mu},\tilde{I}_{2,\lambda}] = \mathcal{O}(\varepsilon^2).
    \end{eqnarray}
    Therefore, upon taking the limit $\varepsilon\to 0$ and comparing the resulting expressions with the algebraic structure of the new group, one finally arrives at
    \begin{equation}
        \label{resultcontra2}
        c_{1,\mu\phantom{2,\lambda}2,\nu}^{\phantom{1,\mu}2,\lambda} = C_{1,\mu\phantom{2,\lambda}2,\nu}^{\phantom{1,\mu}2,\lambda}, \qquad c_{1,\mu\phantom{1,\lambda}2,\nu}^{\phantom{1,\mu}1,\lambda} = 0,
    \end{equation}
    and    
    \begin{equation}
        \label{resultcontra3}
        c_{2,\mu\phantom{1,\lambda}2,\nu}^{\phantom{2,\mu}1,\lambda} = 0, \qquad c_{2,\mu\phantom{2,\lambda}2,\nu}^{\phantom{2,\mu}2,\lambda} = 0,
    \end{equation}
    in that given order.

    Collectively, the set of equalities (\ref{resultcontra1}), (\ref{resultcontra2}), and (\ref{resultcontra3}) accounts for all the distinct aspects involved in the group contraction method. The pair of equations (\ref{resultcontra1}) and (\ref{conditionimportante}) shows that the new Lie group generated by $\{\tilde{I}_{i}\}$ contains a subgroup that is isomorphic to a subgroup of $G$. Finally, while the pair of results (\ref{resultcontra3}) indicates that the set of generators $\{I_{2,\mu}\}$ yields an Abelian subalgebra in the limiting regime, expressions (\ref{resultcontra2}) state that this Abelian subalgebra must be invariant.

    \subsection{A closed $\mathfrak{sim}(2)$ algebra form as a Lorentz limiting case}
    \label{Subsec: 2.B}
    As is well known, the Lorentz algebra is generated by six independent elements, conventionally decomposed into three Lorentz boosts, $(K_1, K_2, K_3)$, and three generators associated with spatial rotations, $(J_1, J_2, J_3)$. As discussed in \cite{primos}, although the emergence of the group $\mathrm{SIM}(2)$ as a contraction of the Lorentz group is guaranteed by the In\"on\"u-Wigner theorem, the explicit construction of a matrix $U$ relating the generators of both algebras is a non-trivial task. To address this issue, it is convenient to begin by organizing the Lorentz generators into a single ordered set,
    \begin{equation}
        I'_i = (K_1, K_2, K_3, J_1, J_2, J_3)^\mathrm{T},\qquad i = 1,2,\dots,6.
    \end{equation}
    One then introduces an automorphism of the Lorentz algebra, represented by a non-singular matrix $\mathcal{A}$, defined through the linear relation
    \begin{equation}
        (T_1, T_2, K_3, J_3, \widetilde{T}_1, \widetilde{T}_2)^\mathrm{T} = \mathcal{A}\cdot(K_1, K_2, K_3, J_1, J_2, J_3)^\mathrm{T},
    \end{equation}
    yielding
    \begin{equation}
        \mathcal{A} = \begin{pmatrix}
    	1 & 0 & 0 & 0 & 1 & 0\\
    	0 & 1 & 0 & -1 & 0 & 0\\
    	0 & 0 & 1 & 0 & 0 & 0\\
    	0 & 0 & 0 & 0 & 0 & 1\\
    	1 & 0 & 0 & 0 & -1 & 0\\
    	0 & 1 & 0 & 1 & 0 & 0
    	\end{pmatrix}.
    \end{equation}
    This transformation amounts to a change of basis in the Lorentz algebra, preserving its structure while recasting it in a form more suitable for the subsequent contraction procedure. The newly introduced generators are given by
    \begin{equation}
        T_1 = K_1 + J_2,\quad T_2 = K_2 - J_1,\quad \widetilde{T}_1 = K_1 - J_2\quad \text{and} \quad \widetilde{T}_2=K_2 + J_1,
    \end{equation}
    which satisfy the commutation relations displayed in Table \ref{Table algebra}. We emphasize that the content of Table \ref{Table algebra} is obtained straightforwardly from the usual commutation relations of the Lorentz algebra, namely,
    \begin{equation}
        \label{LKK}
        [K_i, K_j] = - i \epsilon_{ijk}K_k,
    \end{equation}
    \begin{equation}
        \label{LJK}
        [J_i, K_j] = i \epsilon_{ijk}K_k
    \end{equation}
    and
    \begin{equation}
        \label{LJJ}
        [J_i,J_j] = i\epsilon_{ijk}J_k.
    \end{equation}
    The symbol $\epsilon_{ijk}$ denotes the Levi–Civita tensor, defined such that $\epsilon_{ijk}=+1$ if the ordered triple $(i,j,k)$ corresponds to an even (cyclic) permutation of\footnote{Due to the mostly negative Lorentz metric, from now on, we will apply the prescription $\epsilon_{ijk} \to -\epsilon_{ij}^{\phantom{ij}k}$, so one could thus recover the Einstein summation convention in expressions (\ref{LKK}), (\ref{LJK}), and (\ref{LJJ}).} $(1,2,3)$.
    
    \renewcommand{\arraystretch}{1.4}
    \begin{table}[h!]
        \begin{tabular}{ C{2cm}|C{2cm}|C{2cm}|C{2cm}|C{2cm}|C{2cm}|C{2cm}| }
         $[\text{column},\text{row}]$ & $\mathbf{T_1}$ & $\mathbf{T_2}$ & $\mathbf{K_3}$ & $\mathbf{J_3}$ & $\mathbf{\widetilde{T}_1}$ & $\mathbf{\widetilde{T}_2}$ \\
        \hline
        $\mathbf{T_1}$ & $0$ & $0$ & $iT_1$ & $-iT_2$ & $-2iK_3$ & $-2iJ_3$\\
        \hline
        $\mathbf{T_2}$ & $0$ & $0$ & $iT_2$ & $iT_1$ & $2iJ_3$ & $-2iK_3$ \\
        \hline
        $\mathbf{K_3}$ & $-iT_1$ & $-iT_2$ & $0$ & $0$ & $-i\widetilde{T}_1$ & $-i\widetilde{T}_2$\\
        \hline
        $\mathbf{J_3}$ & $iT_2$ & $-iT_1$ & $0$ & $0$ & $i\widetilde{T}_2$ & $-i\widetilde{T}_1$\\
        \hline
        $\mathbf{\widetilde{T}_1}$ & $2iK_3$ & $-2iJ_3$ & $i\widetilde{T}_1$ & $-i\widetilde{T}_2$ & $0$ & $0$ \\
        \hline
        $\mathbf{\widetilde{T}_2}$ & $2iJ_3$ & $2iK_3$ & $i\widetilde{T}_2$ & $i\widetilde{T}_1$ & $0$ & $0$ \\
        \hline
    \end{tabular}
    \caption{Complete set of commutation relations among the transformed Lorentz algebra generators.}
    \label{Table algebra}
    \end{table}
    \noindent
    
    The development of an approach complementary to the procedure presented in \cite{primos} begins with the attempt to arrange the algebra summarized in Table \ref{Table algebra} into a three-dimensional closed form, mimicking the general aspects of the usual Lorentz algebra given by (\ref{LKK})–(\ref{LJJ}). To this end, we introduce the following triplets of generators:
    \begin{equation}
        \label{novosgen}
        \overline{K}_i = (T_1, T_2, K_3),\quad \overline{J}_i = (0,0,J_3), \quad \text{and} \quad \overline{T}_i = (\widetilde{T}_1, \widetilde{T}_2, 0),
    \end{equation}
    where $i = 1,2,3$. The motivation for, as well as the implications of, this particular arrangement will be discussed at the end of this subsection. Determining the structure constants associated with all possible commutation relations among the generators $\overline{K}_i$, $\overline{J}_i$, and $\overline{T}_i$ is a technically subtle task, which we address below. First, note that these generators can be expressed in terms of the conventional Lorentz generators as
    \begin{eqnarray}
        \label{1}
        \overline{K}_i &=& K_i + \beta_{i}^{\phantom{i}j}J_j,\\
        \label{2}
        \overline{J}_i &=& \gamma_i^{\phantom{i}j}J_j,\\
        \label{3}
        \overline{T}_i &=& \alpha_i^{\phantom{i}j}K_j - \beta_i^{\phantom{i}j}J_j,
    \end{eqnarray}
    where the specific linear combinations are encoded by the constant matrices $\alpha, \beta,$ and $\gamma$, explicitly given by
    \begin{equation}
        \label{defBGA}
        \beta_{i}^{\phantom{i}j} = \begin{pmatrix}
            0  & 1 & 0 \\
            -1 & 0 & 0 \\
            0 & 0 & 0
        \end{pmatrix}, \quad \gamma_{i}^{\phantom{i}j} = \begin{pmatrix}
            0  & 0 & 0 \\
            0 & 0 & 0 \\
            0 & 0 & 1
        \end{pmatrix} \quad \text{and} \quad \alpha_{i}^{\phantom{i}j} = \begin{pmatrix}
            1 & 0 & 0 \\
            0 & 1 & 0 \\
            0 & 0 & 0
        \end{pmatrix}.
    \end{equation}
    Then, every commutation relation within the set $\{\overline{K}_i, \overline{J}_i,\overline{T}_i\}$ can be expressed in terms of the standard Lorentz algebra and computed with the aid of Eqs. (\ref{LKK})–(\ref{LJJ}). The final step consists in reorganizing the expressions so as to recast them in terms of (\ref{novosgen}). The results obtained are gathered below:
    \begin{align}
        \label{AlgnKK}
        [\overline{K}_i,\overline{K}_j] &= -i(\beta_i^{\phantom{i}l}\delta_j^{\phantom{j}m}+\delta_i^{\phantom{i}l}\beta_j^{\phantom{j}m})\epsilon_{lm}^{\phantom{lm}k}\overline{K}_k,\\
        \label{AlgnJK}
        [\overline{J}_i,\overline{K}_j] &= -i\gamma_i^{\phantom{i}l}\epsilon_{lj}^{\phantom{lj}k}\overline{K}_k,\\
        \label{AlgnJJ}
        [\overline{J}_i ,\overline{J}_j] &= 0,\\
        \label{AlgnTT}
        [\overline{T}_i,\overline{T}_j] &= 0,\\ 
        \label{AlgnTK}
        [\overline{T}_i,\overline{K}_j] &= i(\alpha_i^{\phantom{i}l}\delta_j^{\phantom{j}m}+\beta_i^{\phantom{i}l}\beta_j^{\phantom{j}m})\epsilon_{lm}^{\phantom{lm}k}\overline{J}_k + i(\beta_i^{\phantom{i}l}\delta_j^{\phantom{j}m}-\alpha_i^{\phantom{i}l}\beta_j^{\phantom{j}m})\epsilon_{lm}^{\phantom{lm}k}(\gamma_k^{\phantom{k}n}\overline{K}_n + \overline{T}_k),\\
        \label{AlgnTJ}
        [\overline{T}_i,\overline{J}_j] &= -i\gamma_j^{\phantom{j}m}\epsilon_{im}^{\phantom{im}k}\overline{T}_k.
    \end{align}  
    For the sake of clarity and completeness, the full derivation of (\ref{AlgnKK})–(\ref{AlgnTJ}) is presented in Appendix \ref{App: A}. Of course, all the information contained in Table \ref{Table algebra} is encoded in the set of relations presented above.
        
    An alternative three-dimensional formulation of the Lorentz algebra consists of organizing the boost generators into a vectorial array $K_i = (K_1, K_2, K_3)$, while spatial rotations are encoded in a $3\times 3$ antisymmetric matrix defined by
    \begin{equation}
        \label{rotationmatrix}
        M_{ij} = \epsilon_{ij}^{\phantom{ij}k}J_k. 
    \end{equation}
    he definition (\ref{rotationmatrix}) guarantees the correct interpretation of $M_{ij}$ as the generator of three-dimensional rotations. Motivated by this structure, we propose extending the same representation to the $\mathrm{SIM}(2)$ group by introducing a matrix $\overline{M}_{ij}$, given by
    \begin{eqnarray}
        \label{MHATUDO}
        \overline{M}_{ij} &\equiv& \epsilon_{ij}^{\phantom{ij}k}\overline{J}_k\nonumber\\
        &=& \begin{pmatrix}
            0 & -J_3 & 0 \\
            J_3 & 0 & 0 \\
            0 & 0 & 0
        \end{pmatrix}.
    \end{eqnarray}
    With the aid of the standard Levi-Civita identities $\epsilon_{i}^{\phantom{i}lm}\epsilon_{lm}^{\phantom{lm}j}= -2\delta_{i}^{\phantom{i}j}$ and $\epsilon_{ij}^{\phantom{ab}k}\epsilon_{k}^{\phantom{k}lm}=\delta_{i}^{\phantom{i}m}\delta_{j}^{\phantom{j}l}  -\delta_{i}^{\phantom{i}l} \delta_{j}^{\phantom{j}m}$, the commutation relations (\ref{AlgnKK}) -- (\ref{AlgnTJ}) may be then transcribed to the form
    \begin{align}
        \label{AlgnKK2}
        [\overline{K}_i,\overline{K}_j] &= -i(\beta_i^{\phantom{i}l}\delta_j^{\phantom{j}m}+\delta_i^{\phantom{i}l}\beta_j^{\phantom{j}m})\epsilon_{lm}^{\phantom{lm}k}\overline{K}_k,\\
        \label{AlgnJK2}
        [\overline{M}_{ij},\overline{K}_k] &= -i\epsilon_{ij}^{\phantom{ij}l}\gamma_l^{\phantom{i}m}\epsilon_{mk}^{\phantom{mk}n}\overline{K}_n,\\
        \label{AlgnJJ2}
        [\overline{M}_{ij} ,\overline{M}_{lm}] &= 0,\\
        \label{AlgnTT2}
        [\overline{T}_i,\overline{T}_j] &= 0,\\ 
        \label{AlgnTK2}
        [\overline{T}_i,\overline{K}_j] &= i(\alpha_i^{\phantom{i}l}\delta_j^{\phantom{j}m}+\beta_i^{\phantom{i}l}\beta_j^{\phantom{j}m})\overline{M}_{lm} + i(\beta_i^{\phantom{i}l}\delta_j^{\phantom{j}m}-\alpha_i^{\phantom{i}l}\beta_j^{\phantom{j}m})\epsilon_{lm}^{\phantom{lm}k}(\gamma_k^{\phantom{k}n}\overline{K}_n + \overline{T}_k),\\
        \label{AlgnTJ2}
        [\overline{T}_i,\overline{M}_{jk}] &= -i\epsilon_{jk}^{\phantom{jk}l}\gamma_l^{\phantom{l}m}\epsilon_{im}^{\phantom{im}n}\overline{T}_n,
    \end{align}
    which establishes a new convenient closed basis for the Lorentz algebra. Hence, the new basis, composed by $ \{\overline{K}_i, \overline{M}_{ij}, \overline{T}_i\}$, is equally allowed to span the given isomorphic Lorentz group.    
    
    We are now prepared to address the realization of the $\mathrm{SIM}(2)$ group as a limiting case of the Lorentz group. Following the notation established in Eqs.~(\ref{eitaeita}) and (\ref{eitaeita2}), we proceed by decomposing the new set of generators into two distinct sectors:
    \begin{equation}
        \underbrace{(\overline{K}_i, \overline{M}_{ij})}_{\text{``first sector''}}
        \quad \text{and} \quad
        \underbrace{(\overline{T}_i)}_{\text{``second sector''}}.
    \end{equation}
    The key feature of this particular decomposition is that the first sector is chosen to be closed, thereby satisfying the sole requirement for the contraction procedure. In the contraction limit $\varepsilon\to 0$, a new non-isomorphic algebra emerges, whose commutation laws read
    \begin{align}
        \label{AlgnKKfinal}
        [\widehat{K}_i,\widehat{K}_j] &= -i(\beta_i^{\phantom{i}l}\delta_j^{\phantom{j}m}+\delta_i^{\phantom{i}l}\beta_j^{\phantom{j}m})\epsilon_{lm}^{\phantom{lm}k}\widehat{K}_k,\\
        \label{AlgnJKfinal}
        [\widehat{M}_{ij},\widehat{K}_k] &= -i\epsilon_{ij}^{\phantom{ij}l}\gamma_l^{\phantom{i}m}\epsilon_{mk}^{\phantom{mk}n}\widehat{K}_n,\\
        \label{AlgnJJfinal}
        [\widehat{M}_{ij} ,\widehat{M}_{lm}] &= 0,\\
        \label{AlgnTTfinal}
        [\widehat{T}_i,\widehat{T}_j] &= 0,\\ 
        \label{AlgnTKfinal}
        [\widehat{T}_i,\widehat{K}_j] &=  i(\beta_i^{\phantom{i}l}\delta_j^{\phantom{j}m}-\alpha_i^{\phantom{i}l}\beta_j^{\phantom{j}m})\epsilon_{lm}^{\phantom{lm}k}\widehat{T}_k,\\
        \label{AlgnTJfinal}
        [\widehat{T}_i,\widehat{M}_{jk}] &= -i\epsilon_{jk}^{\phantom{jk}l}\gamma_l^{\phantom{l}m}\epsilon_{im}^{\phantom{im}n}\widehat{T}_n,
    \end{align}
    where the new set of generators is defined by 
    \begin{equation}
        \{\overline{K}_i,\overline{M}_{ij},\overline{T}_i\}\xlongrightarrow[]{\varepsilon \to 0} \{\widehat{K}_i,\widehat{M}_{ij},\widehat{T}_i\}.
    \end{equation}
    In particular, the subset $\{\widehat{T}_i\}$ corresponds to the basis of an Abelian and invariant subalgebra, whereas the subset of group generators $\{\widehat{K}_i, \widehat{M}_{ij}\}$ is responsible for spanning the algebra of a Lorentz-type subgroup. From the explicit definitions of $\widehat{K}_i$ and $\widehat{M}_{ij}$, it follows that this subalgebra is ultimately generated by $\{T_1, T_2, K_3,J_3\}$, which precisely corresponds to the set of generators of the $\mathrm{SIM}(2)$ group. Consequently, Eqs. (\ref{AlgnKKfinal}), (\ref{AlgnJKfinal}), and (\ref{AlgnJJfinal}) fully characterize the $\mathfrak{sim}(2)$ Lie algebra, which is given by
    \begin{align}
        \label{AlgnKKfinalfinal}
        [\widehat{K}_i,\widehat{K}_j] &= -i(\beta_i^{\phantom{i}l}\delta_j^{\phantom{j}m}+\delta_i^{\phantom{i}l}\beta_j^{\phantom{j}m})\epsilon_{lm}^{\phantom{lm}k}\widehat{K}_k,\\
        \label{AlgnJKfinalfinal}
        [\widehat{M}_{ij},\widehat{K}_k] &= -i\epsilon_{ij}^{\phantom{ij}l}\gamma_l^{\phantom{i}m}\epsilon_{mk}^{\phantom{mk}n}\widehat{K}_n,\\
        \label{AlgnJJfinalfinal}
        [\widehat{M}_{ij} ,\widehat{M}_{lm}] &= 0,        
    \end{align}
    in the newly constructed basis. As a final remark, we emphasize that since the subalgebra identified here with the $\mathrm{SIM}(2)$ group is a subgroup of the Lorentz group, one could alternatively perform the contraction procedure described in Ref. \cite{primos} and subsequently recover this convenient representation by an analogous reasoning.

    At this point, it is worth proposing a physically reasonable interpretation of the $\mathrm{SIM}(2)$ framework established thus far. This also provides an opportunity to motivate the \textit{Ansatz} introduced in (\ref{novosgen}). As discussed in Refs.~\cite{VSR, GABRIEL, Salinas}, the subset of generators $\{T_1, T_2, K_3\}$ constitutes a consistent framework for defining the analogue of Lorentz boosts within the $\mathrm{SIM}(2)$ setting. Indeed, as demonstrated in the aforementioned references, such transformations preserve the invariance of the speed of light between inertial frames and reproduce effects analogous to Lorentz time dilation and length contraction. Moreover, this set of generators is capable of yielding a well-defined velocity addition law. Additionally, the $\mathrm{SIM}(2)$ group structure features one more generator, denoted by $J_3$. From a physical standpoint, its interpretation parallels that found in the Lorentz case, but in the present scenario the structure effectively restricts the rotational sector to transformations around the $\hat{z}$-axis only.

    Going further, the extension to the $\mathfrak{isim}(2)$ algebra can be readily accomplished. The $\mathrm{ISIM}(2)$ group is formed from the semi-direct product of the $\mathrm{SIM}(2)$ group and the Abelian translation group in $(3+1)$ dimensions, namely $\mathrm{ISIM}(2) = \mathrm{SIM}(2) \rtimes \mathbb{R}^{3,1}$. Accordingly, the construction of a closed representation of the $\mathfrak{isim}(2)$ algebra follows the same blueprint as that adopted in Subsection \ref{Subsec: 2.B}. We begin by incorporating spacetime translations into the set of generators $\{\widehat{K}_i, \widehat{M}_{ij}\}$, and subsequently determine the explicit form of the structure constants that close the algebra. For this task, we shall employ the commutation relations associated with the non-homogeneous sector of the Poincaré group, given by
    \begin{align}
        \label{nhomoLorAlg}
        [K_i, H] = iP_i, \qquad [J_i,P_j] = i\epsilon_{ijk}P_k, \qquad [K_i,P_j]=i\delta_{ij}H,\nonumber\\
        [J_i, H] = 0,\qquad [H,P_i] = 0, \qquad [H,H]=0=[P_i,P_j]. 
    \end{align}
    With these relations, we define a basis for the $\mathfrak{isim}(2)$ algebra as $\{\widehat{K}_i, \widehat{M}_{ij}, H, P_i\}$, where $H \equiv P_0$ and $P_i \equiv (P_1, P_2, P_3)$ are the generators that span the algebra of spacetime translations. Through a direct calculation (following the general scheme presented in Appendix \ref{App: A}), one obtains
    \begin{align}
        [\widehat{K}_i,H] &= iP_i,\\
        [\widehat{K}_i,P_j] &= -i(\eta_{ij}H + \beta_{i}^{\phantom{i}k}\epsilon_{kj}^{\phantom{kj}l}P_l),\\
        [\widehat{M}_{ij},H] &= 0,\\
        [\widehat{M}_{ij},P_k] &= -i\epsilon_{ij}^{\phantom{ij}l}\gamma_{l}^{\phantom{l}m}\epsilon_{mk}^{\phantom{mk}n}P_n,
    \end{align}
    where $\delta_{ij}\equiv -\eta_{ij}$, with $\eta_{ij} = \mathrm{diag}{(-1,-1,-1)}$ denoting the spatial component of the Lorentz metric. This results corresponds to the set of relations required to characterize the $\mathfrak{isim}(2)$ algebra in the chosen basis.

    \section{$\mathrm{SIM}(2)$ group gravitational gauge model}
    \label{Sec:3}

    Let us start with a short outline of the formalism of soft gauge algebras \cite{soft}. This framework generalizes the concept of standard Lie algebras by replacing constant structure constants, $c^{\phantom{i}k}_{i\phantom{k}j}$, with spacetime-dependent structure functions, $c^{\phantom{i}k}_{i\phantom{k}j}(x)$. The hallmark of this approach is a relaxation of the requirements for representations and algebraic consistency. In the conventional scenario, a set of covariant fields $\phi$ carries a representation of a group (characterized by parameters $\epsilon^i$ and generators $L_i$) if the corresponding Lie algebra
    \begin{equation}
        \label{algebrasoft}
        [\delta_{\epsilon_1},\delta_{\epsilon_2}] = \delta_{\epsilon_3}, \qquad \epsilon_{3}^{k}=\epsilon_{1}^{i}\epsilon_{2}^{j}c^{\phantom{i}k}_{i\phantom{k}j},
    \end{equation}
   is realized through the transformation $\delta_\epsilon\phi = -i\epsilon^i\pi(L_i)\phi$, where $\pi(L_i)$ denotes a set of linear operators acting on $\phi$. Conversely, a representation within a soft algebra is characterized by three essential properties: the variation of a field must realize the algebra, maintain linearity with respect to the field, and exhibit homogeneity in the group parameters. Consequently, a consistent candidate for the generator of the representation, $\delta(L_i)$, is defined by the relation
    \begin{equation}
        \label{softger}
        \delta(L_i)\equiv \frac{\partial}{\partial \epsilon^i}\delta_\epsilon\Bigg|_{\epsilon = 0}.
    \end{equation}
    Through the identification $\pi(L_i)\to  \pi(L_i(x))\equiv i \delta(L_i)$, the algebraic relation (\ref{algebrasoft}) is recast as
    \begin{equation}
        [\delta(L_i(x)),\delta(L_j(x))] = -c_{i\phantom{k}j}^{\phantom{i}k}(x)\delta(L_k(x)). 
    \end{equation}
    By construction, the structure fields of the algebra generated by (\ref{softger}) are neither constant ($\partial_\mu c^{\phantom{i}k}_{i\phantom{k}j}(x)\neq 0$) nor invariant ($\delta_\epsilon c^{\phantom{i}k}_{i\phantom{k}j}(x)\neq 0$). Therefore, the standard Jacobi identity gives way to a generalized cyclic identity
    \begin{equation}
        \epsilon^j_3\epsilon^i_2\left(\delta_{\epsilon_1}c^{\phantom{i}k}_{i\phantom{k}j}+\epsilon^l_1c^{\phantom{l}m}_{l\phantom{m}i}c^{\phantom{m}k}_{m\phantom{k}j}\right) + \epsilon^j_1\epsilon^i_3\left(\delta_{\epsilon_2}c^{\phantom{i}k}_{i\phantom{k}j}+\epsilon^l_2c^{\phantom{l}m}_{l\phantom{m}i}c^{\phantom{m}k}_{m\phantom{k}j}\right)+\epsilon^j_2\epsilon^i_1\left(\delta_{\epsilon_3}c^{\phantom{i}k}_{i\phantom{k}j}+\epsilon^l_3c^{\phantom{l}m}_{l\phantom{m}i}c^{\phantom{m}k}_{m\phantom{k}j}\right)=0,
    \end{equation}
    which explicitly accounts for the gauge variations of the structure fields.

    In the limit where the group parameters become spacetime functions ($\epsilon^i\to \epsilon^i(x)$), it becomes necessary to introduce gauge fields, $\Omega^i_{\phantom{i}\mu}(x)$, with transformation laws given by
    \begin{equation}
        \label{vargfield}
        \delta_{\epsilon}\Omega^i_{\phantom{i}\mu} = \partial_{\mu}\epsilon^{i} + \Omega^{k}_{\phantom{j}\mu}\epsilon^{j}c^{\phantom{i}i}_{j\phantom{i}k}(x)\,.
    \end{equation}
    As demonstrated in \cite{soft}, the existence of a covariant derivative $\nabla_\mu = \partial_\mu - \Omega^i_{\phantom{i}\mu}\delta(L_i)$ obeying the property
    \begin{equation}
        \label{relaC11}
        \nabla_\mu(\delta(L_i)\phi) =\delta(L_i)(\nabla_\mu\phi),
    \end{equation}
    so as to preserve the representation of the algebra, is strictly conditional on a fundamental constraint imposed on the structure fields. Specifically, they must be covariantly constant, satisfying $\nabla_\mu c^{\phantom{i}k}_{i\phantom{k}j} = \partial_\mu c^{\phantom{i}k}_{i\phantom{k}j}(x)-\Omega^l_{\phantom{l}\mu}\delta(L_l)c^{\phantom{i}k}_{i\phantom{k}j}(x) \equiv 0$.
    
    The curvature associated with the gauge connection is defined by
    \begin{equation}
        \label{curvgauge}
        \mathbb{R}^{i}_{\phantom{i}\mu\nu} = \partial_{\mu}\Omega^{i}_{\phantom{i}\nu} - \partial_{\nu}\Omega^{i}_{\phantom{i}\mu} + \Omega^k_{\phantom{k}\mu}\Omega^j_{\phantom{j}\nu}c^{\phantom{j}i}_{j\phantom{i}k}(x)\,,
    \end{equation}
    satisfying the generalized Ricci and Bianchi identities, given respectively by
    \begin{equation}
        \label{RicBia}
        [\nabla_{\mu},\nabla_{\nu}] = -\mathbb{R}^{i}_{\phantom{i}\mu\nu}\delta(L_{i})\,, \qquad
        \nabla_{\mu}\mathbb{R}^i_{\phantom{i}\nu\lambda} + \nabla_{\lambda}\mathbb{R}^i_{\phantom{i}\mu\nu} + \nabla_{\nu}\mathbb{R}^i_{\phantom{i}\lambda\mu}=0\,.
    \end{equation}
    Invoking the mixed Jacobi identity involving two covariant derivatives and an arbitrary gauge generator,
    \begin{equation}
        [\delta(L_i),[\nabla_\mu,\nabla_\nu]]+[\nabla_\nu,[\delta(L_i),\nabla_\mu]] + [\nabla_\mu,[\nabla_\nu,\delta(L_i)]]=0,
    \end{equation}
    one verifies that $[\delta(L_i),-\mathbb{R}^{j}_{\phantom{i}\mu\nu}\delta(L_j)]=0$, yielding $\mathbb{R}^{j}_{\phantom{j}\mu\nu}c^{\phantom{j}k}_{j\phantom{k}i}\delta(L_k) = 0$. Since the generators $\delta(L_i)$ are linearly independent, this expression leads directly to the so-called integrability condition
    \begin{equation}
        \label{integrabcond}
        \mathbb{R}^i_{\phantom{i}\mu\nu}=0.
    \end{equation}
    In the next subsection we shall apply this formalism to $\mathrm{ISIM}(2)$ group.
    
    \subsection{Construction of a $\mathrm{ISIM}(2)$ gravitational gauge model}
    
    Consider the $\mathfrak{isim}(2)$ algebra, given by
    \begin{align}
        [\widehat{K}_i,\widehat{K}_j] = (\beta_i^{\phantom{i}l}\delta_j^{\phantom{j}m}+\delta_i^{\phantom{i}l}\beta_j^{\phantom{j}m})\epsilon_{lm}^{\phantom{lm}k}\widehat{K}_k, \quad [\widehat{M}_{ij},\widehat{K}_k] = \epsilon_{ij}^{\phantom{ij}l}\gamma_l^{\phantom{i}m}\epsilon_{mk}^{\phantom{mk}n}\widehat{K}_n,\quad [\widehat{M}_{ij} ,\widehat{M}_{lm}] = 0,\nonumber\\
        [\widehat{K}_i,H] = -P_i,\qquad [\widehat{K}_i,P_j] = (\eta_{ij}H + \beta_{i}^{\phantom{i}k}\epsilon_{kj}^{\phantom{kj}l}P_l), \quad [\widehat{M}_{ij},H] = 0; \quad [\widehat{M}_{ij},P_k] = \epsilon_{ij}^{\phantom{ij}l}\gamma_{l}^{\phantom{l}m}\epsilon_{mk}^{\phantom{mk}n}P_n,
    \end{align}
    where we have conveniently redefined the algebra generators to suppress the imaginary unit. A typical transformation of a covariant field will then be given by
    \begin{equation}
        \delta\phi = \zeta\delta(H)\phi+\zeta^a\delta(P_a)\phi+\lambda^i\delta(\widehat{K}_i)\phi + \frac{1}{2}\lambda^{ij}\delta(\widehat{M}_{ij})\phi,\qquad a,i,j=1,2,3.
    \end{equation}
   The set $\{\zeta, \zeta^a, \lambda^i, \lambda^{ij}\}$ denotes the spacetime-dependent parameters associated with each sector of the gauge transformation. Furthermore, the localization of the symmetry group demands the introduction of a new set of gauge fields, defined respectively by
    \begin{equation}
        \Omega^i_{\phantom{i}\mu}\delta(L_i) = t_\mu\delta(H)+e^a_{\phantom{a}\mu}\delta(P_a) + \omega^i_{\phantom{i}\mu}\delta(\widehat{K}_i)+\frac{1}{2}\omega^{ij}_{\phantom{ij}\mu}\delta(\widehat{M}_{ij}),\quad \mu,\nu, ... =0,1,2,3.
    \end{equation}
    
    Our approach consists of preserving the homogeneous sector intact, spanned by $\{\widehat{K}_i,\widehat{M}_{ij}\}$, while allowing the non-homogeneous sector, spanned by $\{H, P_a\}$, to be softened. Therefore, we must also account for new fields within the theory. For instance, introducing the bookkeeping notation in
    \begin{equation}
        P_A\equiv(H,P_a), \quad \zeta^A\equiv (\zeta,\zeta^a),\quad E^A_{\phantom{A}\mu}\equiv (t_\mu, e^a_{\phantom{a}\mu}),  \qquad A,B,C,...=0,1,2,3\,,
    \end{equation}
    where the initial uppercase Latin letters are reserved for the translational sector of the algebra, we introduce a new set of structure functions through the following commutation relation
    \begin{equation}
        [P_A,P_B]\equiv T_{AB}(H)\delta(H)+T^a_{\phantom{C}AB}(P)\delta(P_a)+R^i_{\phantom{i}AB}(\widehat{K})\delta(\widehat{K}_i)+R^{ij}_{\phantom{ij}AB}(\widehat{M})\delta(\widehat{M}_{ij}). 
    \end{equation}
    It is worth emphasizing that the structure functions $\{T_{AB}(H), T^a_{\phantom{a}AB}(P), R^i_{\phantom{i}AB}(\widehat{K}), R^{ij}_{\phantom{ij}AB}(\widehat{M})\}$ are required to be neither constant nor gauge invariant.

    From (\ref{vargfield}), the transformation laws attended by the gauge fields are given by
    \begin{eqnarray}
        \label{gaugetrans1}
        \delta t_\mu &=& E^C_{\phantom{c}\mu}\zeta^B T_{BC}(H)+\partial_\mu \zeta-\zeta^a\omega_{a\mu} + \lambda^ae_{a\mu},\\
        \label{gaugetrans2}
        \delta e^a_{\phantom{a}\mu} &=& E^C_{\phantom{c}\mu}\zeta^B T^a_{\phantom{a}BC}(P)+\partial_\mu \zeta^a + \zeta \omega^a_{\phantom{a}\mu} + \zeta^d\epsilon_{dc}{}^a\left(\beta_{b}^{\phantom{b}c}\omega^b{}_\mu+\frac{1}{2}\gamma_b{}^c\epsilon_{kl}{}^b\omega^{kl}{}_\mu\right)-\lambda^b(\delta_b{}^at_\mu - \beta_b{}^c\epsilon_{cd}{}^ae^d{}_\mu) -\nonumber\\
        &-& \frac{1}{2}\lambda^{kl}\epsilon_{bd}{}^a\gamma_c{}^d\epsilon_{kl}{}^ce^b{}_\mu,\\
        \label{gaugetrans3}
        \delta\omega^a_{\phantom{a}\mu} &=& E^C_{\phantom{c}\mu}\zeta^B R^a_{\phantom{a}BC}(\widehat{K})+\partial_\mu \lambda^a+\lambda^d\left(\beta_d{}^c\epsilon_{cb}{}^a\omega^b{}_\mu -\beta_b{}^c\epsilon_{cd}{}^a\omega^b{}_\mu-\frac{1}{2}\epsilon_{kl}{}^b\gamma_b{}^c\epsilon_{cd}{}^a\omega^{kl}{}_\mu\right)+ \nonumber\\ &+&\frac{1}{2}\lambda^{kl}\epsilon_{kl}{}^b\gamma_b{}^c\epsilon_{cd}{}^a\omega^d{}_\mu,\\
        \label{gaugetrans4}
        \delta \omega^{ab}{}_\mu &=&  E^C_{\phantom{c}\mu}\zeta^B R^{ab}_{\phantom{a}BC}(\widehat{M})+\frac{1}{2}\partial_\mu \lambda^{ab}.
    \end{eqnarray}
    In the same spirit, definition (\ref{curvgauge}) can be used for the calculation of the total curvatures, which with the aid of the integrability condition (\ref{integrabcond}) yields
    \begin{eqnarray}
        \label{GC1}
        R_{\mu\nu}(H) &=& E^A_{\phantom{A}\mu}E^B_{\phantom{B}\nu}T_{AB}(H),\\
        \label{GC2}
        R^a{}_{\mu\nu}(P) &=& E^A_{\phantom{A}\mu}E^B_{\phantom{B}\nu}T^a_{\phantom{a}AB}(P),\\
        \label{GC3}
        R^i{}_{\mu\nu}(\widehat{K}) &=& E^A_{\phantom{A}\mu}E^B_{\phantom{B}\nu}R^i_{\phantom{a}AB}(\widehat{K}),\\
        \label{GC4}
        R^{ij}{}_{\mu\nu}(\widehat{M}) &=& E^A_{\phantom{A}\mu}E^B_{\phantom{B}\nu}R^{ij}_{\phantom{a}AB}(\widehat{M}),
    \end{eqnarray}
    where we had defined the gauge curvatures as
    \begin{eqnarray}
        \label{curvH}
        R_{\mu\nu}(H) &=& \partial_\mu t_\nu - \partial_\nu t_\mu + e_{a\mu}\omega^a{}_\nu - e_{a\nu}\omega^a{}_\mu,\\
        \label{curvP}
        R^a{}_{\mu\nu}(P) &=& \partial_\mu e^a{}_\nu - \partial_\nu e^a{}_\mu -t_\mu \omega^a{}_\nu + t_\nu \omega^a{}_\mu+\beta_b{}^c\epsilon_{cd}{}^a(e^d{}_\mu\omega^b{}_\nu - e^d{}_\nu\omega^b{}_\mu)+\nonumber\\ &+&\frac{1}{2}\epsilon_{kl}{}^c\gamma_c{}^d\epsilon_{db}{}^a(e^b{}_\mu \omega^{kl}{}_\nu - e^b{}_\nu\omega^{kl}{}_\mu),\\
        \label{curvK}
        R^a{}_{\mu\nu}(\widehat{K}) &=& \partial_\mu \omega^a{}_\nu - \partial_\nu \omega^a{}_\mu-\beta_b{}^c\epsilon_{cd}{}^a(\omega^b{}_\mu\omega^d{}_\nu-\omega^b{}_\nu\omega^d{}_\mu)+\frac{1}{2}\epsilon_{kl}{}^b\gamma_b{}^c\epsilon_{cd}{}^a(\omega^d{}_\mu\omega^{kl}{}_\nu-\omega^d{}_\nu\omega^{kl}{}_\mu),\\
        \label{curvM}
        R^{ab}_{\phantom{ab}\mu\nu}(\widehat{M}) &=& \frac{1}{2}\left(\partial_\mu \omega^{ab}{}_\nu-\partial_\nu \omega^{ab}{}_\mu\right).
    \end{eqnarray}
    Now it is possible to properly discuss the interpretation of the fields $E^A{}_\mu(x)$. Since the $\mathrm{ISIM}(2)$ group is obtained as a well-defined contraction of the Poincaré group, where the translational sector remains intact, one is strongly inclined to interpret $E^A{}_\mu(x)$ as the vielbein, with $t_\mu$ and $e^a{}_\mu$ being its respective components. Notably, the sole fields whose transformation laws are affected by translations are those comprising $E^A{}_\mu(x)$. In analogy with \cite{Andringa2}, this fact becomes particularly relevant when considering the torsionless regime, wherein $t_\mu$ and $e^a{}_\mu$ can be regarded as independent fields.

    Following the construction presented in \cite{Hartong}, we propose a frame bundle structure $(\mathrm{FM}, \pi, M)$, with the total space defined by the disjoint union of all ordered bases over the manifold $M$, given by  
    \begin{equation}
        \mathrm{FM} = \bigcup_{p \in M}(p,\{E^A(p)\}),
    \end{equation}
   where $\{E^A(p)\}$ denotes the set of all possible bases associated with the tangent space at the point $p\in M$, and a usual canonical projection map is assumed. In this scenario, the structure group is chosen to be the homogeneous part of the $\mathrm{VSR}$ symmetry group, consisting of the combination of $\mathrm{VSR}$ boosts and rotations. Upon fixing the frame components as $E^A=(t,e^a)$ and introducing the internal metric tensor $\eta_{AB}$
    \begin{equation}
        \eta_{AB} = \begin{pmatrix}
            1& 0 \\
            0& \eta_{ab}
        \end{pmatrix},
    \end{equation}
    defined blockwise such that the temporal component is unity and the spatial sector is given by $\eta_{ab}\equiv \mathrm{diag}(-1,-1,-1),$ one immediately verifies that the co-frame field satisfies
    \begin{equation}
        E_A \equiv (t,e_a) = (t,\eta_{ab}e^b).
    \end{equation}
    Furthermore, by means of the tetrad field components, the first orthogonality relation, $E^A{}_\mu E_B{}^\mu = \delta^A{}_B$, implies that 
    \begin{eqnarray}
        t^\mu t_\mu = 1,\qquad e^a{}_\mu t^\mu = 0, \qquad  e_a{}^\mu t_\mu=0, \qquad \text{and} \qquad e^a{}_\mu e _b{}^\mu = \delta^a{}_b,
    \end{eqnarray}
    whereas the second orthogonality relation, $E^A{}_\mu E_A{}^\nu = \delta^\nu{}_\mu$, yields
    \begin{eqnarray}
        \label{WHOAHH}
        e^a{}_\mu e_a{}^\nu = \delta^\nu{}_\mu - t^\nu t_\mu. 
    \end{eqnarray}
    Crucially, it then becomes evident from Eqs. (\ref{GC1})–-(\ref{GC4}) that the gauge curvature components manifest themselves on the base manifold through the projection of internal-space objects defined over the fiber. This correspondence offers a clear geometrical interpretation of the structure fields in non-supersymmetric gravitational theories.
    
    The projection of the internal Minkowskian metric to the base manifold through the action of tetrad fields induces a spacetime metric tensor $g_{\mu\nu}$ defined by
    \begin{eqnarray}
        \label{STMET}
        g_{\mu\nu} &\equiv& E^A{}_\mu E^B{}_\nu \eta_{AB}\nonumber\\
        &=& t_\mu t_\nu + e^a{}_\mu e_{a\nu}.
    \end{eqnarray}
    By introducing a spacetime-dependent parameter $\xi^\mu$ such that
    \begin{equation}
        \label{projxi}
        \zeta^A= E^A{}_\mu \xi^\mu,\qquad \lambda^a = \omega^a{}_\mu \xi^\mu, \qquad \text{and} \qquad \lambda^{ab} = \omega^{ab}{}_\mu \xi^\mu,
    \end{equation}
    it can be shown that each gauge transformation (\ref{gaugetrans1}) -- (\ref{gaugetrans4}) reduces to a Lie derivative parametrized by $\xi^\mu$: 
    \begin{eqnarray}
        \label{tandandandandandnadna1}
        \delta t_\mu &=& \xi^\nu \partial_\nu t_\mu + \partial_\mu \xi^\nu t_\nu\\
        \label{tandandandandandnadna2}
        \delta e^a{}_\mu &=& \xi^\nu \partial_\nu e^a{}_\mu + \partial_\mu \xi^\nu e^a{}_\nu\\ 
        \delta\omega^a{}_\mu &=& \xi^\nu \partial_\nu \omega^a{}_\mu + \partial_\mu \xi^\nu \omega^a{}_\nu\\
        \label{tandandandandandnadna4}
        \delta\omega^{ab}{}_\mu &=& \frac{1}{2}\left(\xi^\nu \partial_\nu \omega^{ab}{}_\mu + \partial_\mu \xi^\nu \omega^{ab}{}_\nu\right).        
    \end{eqnarray}
    Ultimately, this property justifies the explicit role of the structure fields within the transformation laws of the gauge fields: in tandem with the definition of $\xi^\mu$, they also provide a curvature correction that effectively bridges gauge transformations and spacetime diffeomorphisms. Equation (\ref{STMET}) yields
    \begin{equation}
        \delta g_{\mu\nu} = \xi^\alpha \partial_\alpha g_{\mu\nu} + \partial_\mu \xi^\alpha g_{\alpha \nu} + \partial_\nu \xi^\alpha g_{\mu\alpha},
    \end{equation}
    demonstrating that $g_{\mu\nu}$ transforms as a symmetric $(0,2)$-tensor under Lie derivatives. For completeness, applying the orthogonality relations to the first expression in (\ref{projxi}) yields the explicit form of $\xi^\mu$ in terms of the independent gauge fields $t_\mu$ and $e^a{}_\mu$: $\xi^\mu = \zeta t^\mu + \zeta^a e_a{}^\mu$.
   
    Let $V^\mu = E_A{}^\mu V^A$ be an arbitrary vector field defined over the base space manifold. The analysis of the vielbein postulate is established by imposing the identity
    \begin{equation}
        \label{tetradpostulate}
        D_\mu V^A = E^A{}_\nu\widehat{D}_\mu V^\nu.
    \end{equation}
    The object $D_\mu$, explicitly given by
    \begin{equation}
        D_\mu \equiv \partial_\mu - \omega^i{}_\mu\delta(\widehat{K}_i)-\frac{1}{2}\omega^{ij}{}_\mu\delta(\widehat{M}_{ij}),
    \end{equation}
    denotes the covariant derivative with respect to the $\mathrm{SIM}(2)$ group structure. Conversely, $\widehat{D}_\mu$, expressed as  
    \begin{equation}
        \widehat{D}_\mu V^\nu = \partial_\mu V^\nu + \widehat{\Gamma}^\nu{}_{\mu\alpha}V^\alpha,
    \end{equation}
    stands for the covariant derivative acting on the base space, defined in terms of an affine connection $\widehat{\Gamma}^\alpha{}_{\mu\nu}$. The content of Eq. (\ref{tetradpostulate}) can be decomposed into two other equations,
    \begin{eqnarray}
        \label{anima1}
        D_\mu(t_\nu V^\nu) &=& t_\nu \widehat{D}_\mu V^\nu,\\
        \label{anima2}
        D_\mu(e^a{}_\nu V^\nu) &=& e^a{}_\nu \widehat{D}_\mu V^\nu,
    \end{eqnarray}
    which will be inspected sequentially. We shall examine the expression (\ref{anima1}) in detail, from which the development of (\ref{anima2}) follows directly. Noting that $D_\mu V^\nu = \partial_\mu V^\nu$, Eq. (\ref{anima1}) expands as 
    \begin{equation}
        \label{TP!}
        \left(\partial_\mu t_\nu -  \omega^i{}_\mu\delta(\widehat{K}_i)t_\nu -\frac{1}{2}\omega^{ij}{}_\mu\delta(\widehat{M}_{ij})t_\nu - t_\rho \widehat{\Gamma}^\rho{}_{\mu\nu}\right)V^\nu =0. 
    \end{equation}
    Based on (\ref{gaugetrans1}), the $\mathrm{VSR}$-boosts transformation of the gauge field $t_\mu$ is identified as
    \begin{equation}
        \lambda^i\delta(\widehat{K}_i)t_\nu = -\lambda^ie_{i\nu}
    \end{equation}
    and by virtue of Eq. (\ref{projxi}), the expression above yields an explicit form for $\omega^i{}_\mu \delta(\widehat{K}_i)t_\nu$
    \begin{equation}
        \label{adaslav1}
        \omega^i{}_\mu \delta(\widehat{K}_i)t_\nu=\omega^i{}_\mu  e_{i\nu}.
    \end{equation}
    Exploiting the fact that $t_\mu$ is invariant under $\mathrm{VSR}$-rotations, one can readily show that
    \begin{equation}
        \label{adaslav2}
        \frac{1}{2}\omega^{ij}{}_\mu \delta(\widehat{M}_{ij})t_\nu = 0.
    \end{equation}
    Hence, the employment of (\ref{adaslav1}) and (\ref{adaslav2}) into (\ref{TP!}), gives the first component of the vielbein postulate
    \begin{equation}
        \label{VBP1}
        \partial_\mu t_\nu - \omega^a{}_\mu e_{a\nu} - \widehat{\Gamma}^\rho{}_{\mu\nu}t_\rho = 0.
    \end{equation}
    
    Applying an analogous reasoning to equation (\ref{anima2}) we have
    \begin{equation}
        \label{VBP2}
        \partial_\mu e^a{}_\nu + \omega^b{}_\mu(\delta_b{}^at_\nu-\beta_b{}^c\epsilon_{cd}{}^ae^d{}_\nu)-\frac{1}{2}\omega^{kl}{}_\mu\epsilon_{kl}{}^c\gamma_c{}^d\epsilon_{db}{}^ae^b{}_\nu-\widehat{\Gamma}^\rho{}_{\mu\nu}e^a{}_\rho = 0.
    \end{equation}
    Combining Eqs. (\ref{VBP1}) and (\ref{VBP2}) enables a solution for the affine connection $\widehat{\Gamma}^\rho{}_{\mu\nu}$. Specifically, after contracting Eq. (\ref{VBP2}) with $e_a{}^\sigma$, utilizing relation (\ref{WHOAHH}), and introducing Eq. (\ref{VBP1}), one arrives at
    \begin{equation}
        \label{gamminha maluco}
        \widehat{\Gamma}^\sigma{}_{\mu\nu}= t^\sigma\left(\partial_\mu t_\nu -\omega^a{}_\mu e_{a\nu}\right)+e_a{}^\sigma\left[\partial_\mu e^a{}_\nu + \omega^b{}_\mu(\delta_b{}^at_\nu-\beta_b{}^c\epsilon_{cd}{}^ae^d{}_\nu)- \frac{1}{2}\omega^{kl}{}_\mu\epsilon_{kl}{}^c\gamma_c{}^d\epsilon_{db}{}^ae^b{}_\nu\right].
    \end{equation}
    A straightforward computation of the antisymmetric part of (\ref{gamminha maluco}) results in\footnote{Here, we adopt the standard anti-symmetrization convention $A_{[\mu\nu]}=\frac{1}{2}(A_{\mu\nu}-A_{\nu\mu})$.}
    \begin{equation}
        2\widehat{\Gamma}^\sigma{}_{[\mu\nu]}= t^\sigma R_{\mu\nu}(H) + e_a{}^\sigma R^a{}_{\mu\nu}(P),
    \end{equation}
    implying that the spacetime torsion tensor can be expressed as a linear combination of the curvatures associated with the non-homogeneous sector of the $\mathfrak{isim}(2)$ algebra. 
    
    From this point forward, we restrict our analysis to a torsionless scenario, characterized by the imposition of a zero-torsion condition on the fiber (see, e.g., Section 7.3 of \cite{piorartigodomundo}). In the present framework, this geometric restriction translates into the so-called conventional curvature constraints: $T_{AB}(H) = 0 = T^a{}_{AB}(P)$. While this is not the most general case, its consideration allows for an analytic action built upon geometric tools completely determined in terms of fields with a clear physical interpretation. The imposition of the metric compatibility condition over the spacetime metric (\ref{STMET}), explicitly
    \begin{equation}
        \widehat{D}_\alpha g_{\mu\nu} = \partial_\alpha g_{\mu\nu} - \widehat{\Gamma}^\lambda{}_{\alpha\mu}g_{\lambda \nu} - \widehat{\Gamma}^\lambda{}_{\alpha\nu}g_{\mu\lambda} \equiv 0,
    \end{equation}    
    furnishes, as usual, a unique solution for $\widehat{\Gamma}^\sigma{}_{\mu\nu}$ given by
    \begin{equation}
        \label{gamminha maluco2}
        \widehat{\Gamma}^\sigma{}_{\mu\nu}=\frac{1}{2}g^{\sigma\lambda}\left(\partial_\mu g_{\lambda \nu} + \partial_\nu g_{\lambda\mu}-\partial_\lambda g_{\mu\nu}\right).
    \end{equation}
    An important feature of the torsionless scenario becomes evident from the compatibility between the connections. To verify this consistency, one begins by expanding Eq. (\ref{gamminha maluco2}) with the aid of (\ref{STMET}), and then introducing (\ref{curvH}) and (\ref{curvP}) under the conventional curvature constraints.
    
    Another feature to be explored is that, in close analogy with the standard formulation of general relativity, these conventional constraints allow one to solve the algebraic equations for the gauge fields $\omega^a{}_\mu$ and $\omega^{ij}{}_\mu$, thereby rendering them dependent fields in terms of the fundamental dynamical variables $t_\mu$ and $e^a{}_\mu$. Specifically, from $R^a{}_{\mu\nu}(P)=0$, one obtains
    \begin{equation}
        -e^b{}_\mu \widetilde{\omega}_b{}^a{}_\nu + e^b{}_\nu\widetilde{\omega}_b{}^a{}_\mu = \partial_\mu e^a{}_\nu - \partial_\nu e^a{}_\mu -t_\mu \omega^a{}_\nu + t_\nu \omega^a{}_\mu+\beta_b{}^c\epsilon_{cd}{}^a(e^d{}_\mu\omega^b{}_\nu - e^d{}_\nu\omega^b{}_\mu),
    \end{equation}
    where $\widetilde{\omega}_b{}^a{}_\mu=-\widetilde{\omega}^a{}_{b\nu}$ is an antisymmetric field defined as 
    \begin{equation}
        \widetilde{\omega}_b{}^a{}_\mu=\frac{1}{2}\omega^{kl}{}_\mu\epsilon_{kl}{}^c\gamma_c{}^d\epsilon_{db}{}^a.
    \end{equation}
    Owing to the properties of the Levi-Civita symbol and the definition of $\gamma_a{}^b$, the only non-vanishing components of $\widetilde{\omega}_b{}^a{}_\mu$ are those corresponding to $\omega^{12}{}_\mu$ and $\omega^{21}{}_\mu$, which fully encompass the degrees of freedom of the $\mathrm{VSR}$ rotational symmetry sector. By introducing the definition $\mathcal{K}^a{}_{\mu\nu} = -e^b{}_\mu \widetilde{\omega}_b{}^a{}_\nu + e^b{}_\nu\widetilde{\omega}_b{}^a{}_\mu$, one shows that explicitly evaluating the cyclic-like permutation $e^{i\nu}e^{j\rho}(\mathcal{K}^a{}_{\mu\nu}e_{a\rho}+\mathcal{K}^a{}_{\rho\mu}e_{a\nu}-\mathcal{K}^a{}_{\nu\rho}e_{a\mu})$ yields the solution for $\widetilde{\omega}^{ij}{}_\mu$
    \begin{eqnarray}
        \label{solrot}
        \widetilde{\omega}^{ij}{}_\mu = 2e^{[i\alpha}\partial_{[\mu}e^{j]}{}_{\alpha]}-t_\mu e^{[i}{}_\alpha\omega^{j]\alpha} - e^{i\alpha}e^{j\beta}e_{a\mu}\partial_{[\alpha}e^a{}_{\beta]}+ 2e^{[i\alpha}\beta_b{}^c\epsilon_{cd}{}^{j]}e^d{}_{[\mu}\omega^b{}_{\alpha]} - e^{i\alpha}e^{j\beta}e_{a\mu}\beta_b{}^c\epsilon_{cd}{}^a e^d{}_{[\alpha}\omega^b{}_{\beta]}.
    \end{eqnarray}
    The search for a solution for $\omega^a{}_\mu$ starts from the computation of the left-hand side of $e^{i\mu}t^\nu R^a{}_{\mu\nu}(P)=0$
    \begin{equation}
        e^{i\mu}t^\nu(\partial_\mu e^a{}_\nu-\partial_\nu e^a{}_\mu)+e^{i\mu}\omega^a{}_\mu +\beta_b{}^c\epsilon_{c}{}^{ia}t^\nu\omega^b{}_\nu+t^\nu \widetilde{\omega}^{ia}{}_\nu = 0. 
    \end{equation}
    Then, after introducing (\ref{solrot}) into the above expression, one arrives at
    \begin{equation}
        \label{intriaiosdldp´sl}
        e^{i\mu}t^\nu(\partial_\mu e^a{}_\nu-\partial_\nu e^a{}_\mu)+e^{i\mu}\omega^a{}_\mu+2t^\nu e^{[i\alpha}\partial_{[\nu}e^{a]}{}_{\alpha]}-\frac{1}{2}(e^{i\alpha}\omega^a{}_\alpha-e^{a\alpha}\omega^i{}_\alpha)=0.
    \end{equation}
    However, from $e^{a\mu}e^{i\nu}R_{\mu\nu}(H) = 0 $, one finds that
    \begin{equation}
        -(e^{i\alpha}\omega^a{}_\alpha-e^{a\alpha}\omega^i{}_\alpha)=e^{a\mu}e^{i\nu}(\partial_\mu t_\nu- \partial_\nu t_\mu)
    \end{equation}
    and the expression (\ref{intriaiosdldp´sl}) becomes
    \begin{equation}
        \label{picoo}
        e^{i\mu}t^\nu(\partial_\mu e^a{}_\nu-\partial_\nu e^a{}_\mu)+e^{i\mu}\omega^a{}_\mu+2t^\nu e^{[i\alpha}\partial_{[\nu}e^{a]}{}_{\alpha]}+\frac{1}{2}e^{a\mu}e^{i\nu}(\partial_\mu t_\nu- \partial_\nu t_\mu)=0.
    \end{equation}
    After a convenient manipulation of indices Eq. (\ref{picoo}) gives
    \begin{equation}
        e^{i\mu}\omega^a{}_\mu =  e^{i\mu}\left[t^\nu e^{a\alpha}e^c{}_\mu \partial_{[\alpha}e_{c\nu]}-t^\nu\partial_{[\mu}e^a{}_{\nu]}+e^{a\nu}\partial_{[\mu}t_{\nu]}\right],
    \end{equation}
    finally yielding to the desired result
    \begin{equation}
        \label{solboo}
        \omega^a{}_\mu = t^\nu e^{a\alpha}e^c{}_\mu \partial_{[\alpha}e_{c\nu]}-t^\nu\partial_{[\mu}e^a{}_{\nu]}+e^{a\nu}\partial_{[\mu}t_{\nu]}.
    \end{equation}
    As a consequence of (\ref{solrot}) and (\ref{solboo}) we are left with $e^a{}_\mu$ and $t_\mu$ as independent fields of the theory.

   The last ingredient in this gravity model is the investigation of the spacetime curvature, denoted by $\widehat{R}^\sigma{}_{\rho\mu\nu}$, which is obtained from the commutator $[\widehat{D}_\mu, \widehat{D}_\nu]V^\sigma = \widehat{R}^\sigma{}_{\rho\mu\nu} V^\rho$ as follows
    \begin{equation}
        \label{STcurv}
        \widehat{R}^\sigma{}_{\rho\mu\nu} = \partial_\mu \widehat{\Gamma}^\sigma{}_{\rho\nu} - \partial_\nu \widehat{\Gamma}^\sigma{}_{\rho\mu} + \widehat{\Gamma}^\sigma{}_{\lambda\mu}\widehat{\Gamma}^\lambda{}_{\beta\nu} - \widehat{\Gamma}^\sigma{}_{\lambda\nu}\widehat{\Gamma}^\lambda{}_{\beta\mu}. 
    \end{equation}
    As covered in detail in Appendix \ref{App: B}, the introduction of (\ref{gamminha maluco}) into (\ref{STcurv}), results in
    \begin{equation}
        \label{auidsauiddadasdas}
        \widehat{R}^\sigma{}_{\rho\mu\nu}= R^a{}_{\mu\nu}(\widehat{K})t_\rho e^\sigma{}_a - \left(R^c{}_{\mu\nu}(\widehat{K})\beta_c{}^n\epsilon_{nd}{}^a + R^{kl}{}_{\mu\nu}(\widehat{M})\epsilon_{kl}{}^n\gamma_n{}^c\epsilon_{cd}{}^a\right)e^d{}_\rho e_a{}^\sigma.
    \end{equation}
    Therefore, a good candidate to compose a functional action is naturally given by $\widehat{R} = g^{\mu\nu}\widehat{R}_{\mu\nu}$, where $\widehat{R}_{\mu\nu}$, explicitly denoted by $\widehat{R}_{\mu\nu} = \widehat{R}^\alpha{}_{\mu\alpha\nu}$, is a geometric object equivalent to the Ricci tensor in the standard framework of general relativity. In terms of an integral action, it is then natural to propose
    \begin{equation}
        \label{ação integral}
        S^{Grav}_{VSR} = \int d^4x\, \sqrt{-g}\,\widehat{R},
    \end{equation}
    where $g\equiv \det{g_{\mu\nu}}$ is introduced to guarantee that (\ref{ação integral}) remain invariant under diffeomorphisms.
    
    \section{Final remarks}
    
    Crucially, $\mathrm{VSR}$ breaks Lorentz invariance not via external background fields, but by substituting the fundamental symmetry group, thereby implying a privileged light-like direction that allows for a controlled introduction of anisotropies while preserving classical relativistic tests \cite{Anisotropia2}. In this paper, we have obtained a closed-form expression for the $\mathfrak{isim}(2)$ algebra, paving the way for further physical constructions. In particular, we explored the gravitational model arising from gauging the parameters of this algebra in the so-called soft algebra context.
    
    The resulting gravitational model has been explored in detail, and our construction points to the possibility of studying physically relevant gravitational scenarios. By the very nature of the underlying gauged symmetry, we believe that this framework may naturally encompass, for instance, Bianchi cosmological solutions. Furthermore, since the $\mathrm{ISIM}(2)$ symmetry carries a privileged direction, it is pertinent to frame this gravitational model within the context of Finsler geometry \cite{Bogo1}. In particular, while the connection is unique in our torsionless scenario, it would be intriguing to analyze whether a multiple-connection formalism arises, paralleling that presented in \cite{Bogo1}, when torsion is introduced. Some developments presented in \cite{Bogo2} suggest this connection, but a more careful analysis is required. Moreover, the gravitational scenario constructed here opens the possibility of searching for gravitational and cosmological solutions to be contrasted with those previously investigated in the realm of Finsler geometry \cite{GibbonsisFinsler,KouretsisinFinsler}. We shall delve into such possibilities in a forthcoming paper.
    
\section*{Acknowledgment}
\noindent
The authors thank Professor Gustavo P. de Brito for helpful conversations. JMBM thanks Coordenação de Aperfeiçoamento de Pessoal de Nível Superior (CAPES - Finance Code 001) and JMHS thanks CNPq (grant No. 307641/2022-8) for financial support. 

\section*{Declaration of generative AI and AI-assisted technologies in the manuscript preparation process}

During the preparation of this work the authors used Gemini as a language reviewer. After using this tool/service, the authors reviewed and edited the content as needed and take full responsibility for the content of the article.

    \appendix
    \section{Detailed derivation of expressions (\ref{AlgnKK}) to (\ref{AlgnTJ})}
    \label{App: A}

    In this appendix we provide the explicit derivation of the commutation relations listed in Eqs. (\ref{AlgnKK}) -- (\ref{AlgnTJ}), following the same order in which these relations were originally presented. We begin with the commutator $[\overline{K}_i,\overline{K}_j]$. Using the definition of $\overline{K}$ (cf. Eq. (\ref{1})), the Lie bracket expands as
    \begin{align}
        [\overline{K}_i,\overline{K}_j] &= [K_i+\beta_{i}^{\phantom{i}l}J_l,K_j+\beta_{j}^{\phantom{j}m}J_m]\nonumber\\
        &= [K_i,K_j] + \beta_{j}^{\phantom{j}m}[K_i,J_m]+\beta_{i}^{\phantom{i}l}[J_l,K_j] + \beta_{i}^{\phantom{i}l}\beta_{j}^{\phantom{j}m}[J_l,J_m].
    \end{align}
    Invoking the Lorentz algebra relations for $[K,K]$, $[K,J]$ and $[J,J]$, together with the convention specified in the main text, one finds
    \begin{equation}
        \label{app1}
        [\overline{K}_i,\overline{K}_j] = i(\delta_{i}^{\phantom{i}l}\delta_{j}^{\phantom{j}m}-\beta_{i}^{\phantom{i}l}\beta_{j}^{\phantom{j}m})\epsilon_{lm}^{\phantom{lm}k}J_k - i(\beta_{i}^{\phantom{i}l}\delta_{j}^{\phantom{j}m}+\delta_{i}^{\phantom{i}l}\beta_{j}^{\phantom{j}m})\epsilon_{lm}^{\phantom{lm}k}K_k.
    \end{equation}
    According to Table \ref{Table algebra}, commutators between elements of $\overline{K}$ must close on the same set. Therefore, the right-hand side of (\ref{app1}) must be expressible purely in terms of $\overline{K}$. Indeed, direct component-wise comparison shows that
    \begin{equation}
        \label{apppelotinha}
        (\delta_{i}^{\phantom{i}l}\delta_{j}^{\phantom{j}m}-\beta_{i}^{\phantom{i}l}\beta_{j}^{\phantom{j}m})\epsilon_{lm}^{\phantom{lm}k} = -(\beta_{i}^{\phantom{i}l}\delta_{j}^{\phantom{j}m}+\delta_{i}^{\phantom{i}l}\beta_{j}^{\phantom{j}m})\epsilon_{lm}^{\phantom{lm}n}\beta_n^{\phantom{n}k}.
    \end{equation}
    Using the identity (\ref{apppelotinha}), the commutator can be rewritten as
    \begin{equation}
        [\overline{K}_i,\overline{K}_j] = -i(\beta_{i}^{\phantom{i}l}\delta_{j}^{\phantom{j}m}+\delta_{i}^{\phantom{i}l}\beta_{j}^{\phantom{j}m})\epsilon_{lm}^{\phantom{lm}k}(K_k+\beta_{k}^{\phantom{k}n}J_n),
    \end{equation}
    or, taking (\ref{1}) into consideration the definition, as
    \begin{equation}
        \label{levitate}
        [\overline{K}_i,\overline{K}_j] = -i(\beta_{i}^{\phantom{i}l}\delta_{j}^{\phantom{j}m}+\delta_{i}^{\phantom{i}l}\beta_{j}^{\phantom{j}m})\epsilon_{lm}^{\phantom{lm}k}\overline{K}_k,
    \end{equation}
    which establishes Eq. (\ref{AlgnKK}).

    The evaluation of $[\overline{J}_i,\overline{K}_j]$ follows analogous steps that led us to (\ref{levitate}) and will therefore be presented more succinctly. Using the definitions of $\overline{K}$ and $\overline{J}$, and subsequently taking the Lorentz algebra into account, one obtains
    \begin{equation}
        \label{app2}
        [\overline{J}_i,\overline{K}_j] = -\gamma_i^{\phantom{i}l}\epsilon_{lj}^{\phantom{lj}k}K_k - i \gamma_i^{\phantom{i}l}\beta_j^{\phantom{j}m}\epsilon_{lm}^{\phantom{lm}k}J_k.
    \end{equation}
    Again, by the same reasoning, expression (\ref{app2}) must be proportional to $\overline{K}$. Hence, considering the identity
    \begin{equation}
        \gamma_i^{\phantom{i}l}\beta_j^{\phantom{j}m}\epsilon_{lm}^{\phantom{lm}k} = \gamma_i^{\phantom{i}l}\epsilon_{lj}^{\phantom{lj}n}\beta_n^{\phantom{n}k},
    \end{equation}
    expression (\ref{app2}) reduces to
    \begin{equation}
        [\overline{J}_i,\overline{K}_j] = -\gamma_i^{\phantom{i}l}\epsilon_{lj}^{\phantom{lj}k}\overline{K}_k,  
    \end{equation}
    as required. 
    
    The commutator $[\overline{J}_i,\overline{J}_j]$ is straightforward. From the explicit form of $\overline{J}$, one finds
    \begin{equation}
        \label{appformo}
        [\overline{J}_i,\overline{J}_j] = i\gamma_{i}^{\phantom{i}l}\gamma_{j}^{\phantom{j}m}\epsilon_{lm}^{\phantom{lm}k}J_k.
    \end{equation}
    Since the only non-vanishing component of $\gamma_{i}^{\phantom{i}j}$ lies along the third direction, all contributions in (\ref{appformo}) vanish identically, yielding
    \begin{equation}
        [\overline{J}_i,\overline{J}_j] = 0.
    \end{equation}

    We now turn to the commutators involving $\overline{T}_i$. From (\ref{3}), the bracket $[\overline{T}_i,\overline{T}_j]$ becomes
    \begin{equation}
        \label{appTT}
        [\overline{T}_i,\overline{T}_j] = i(\beta_{i}^{\phantom{i}l}\alpha_{j}^{\phantom{j}m}+\alpha_{i}^{\phantom{i}l}\beta_{j}^{\phantom{j}m})\epsilon_{lm}^{\phantom{lm}k}K_k+(\alpha_{i}^{\phantom{i}l}\alpha_{j}^{\phantom{j}m}-\beta_{i}^{\phantom{i}l}\beta_{j}^{\phantom{j}m})\epsilon_{lm}^{\phantom{lm}k}J_k.
    \end{equation}
    A direct inspection using the explicit forms of $\alpha$ and $\beta$ shows that both coefficients vanish for all possible values of $i$ and $j$. Hence,
    \begin{equation}
        [\overline{T}_i,\overline{T}_j] = 0.
    \end{equation}
    
    Next, we evaluate the commutator between the generators $\overline{T}_i$ and $\overline{K}_j$. Using their definition, a direct computation yields
    \begin{equation}
        \label{apptustustus}
        [\overline{T}_i,\overline{K}_j] = i(\beta_{i}^{\phantom{i}l}\delta_{j}^{\phantom{j}m}-\alpha_{i}^{\phantom{i}l}\beta_{j}^{\phantom{j}m})\epsilon_{lm}^{\phantom{lm}k}K_k + i\alpha_{i}^{\phantom{i}l}\delta_{j}^{\phantom{j}m}\epsilon_{lm}^{\phantom{lm}k}J_k+i\beta_{i}^{\phantom{i}l}\beta_{j}^{\phantom{j}m}\epsilon_{lm}^{\phantom{lm}k}J_k.
    \end{equation}
    From the definition of the matrices $\alpha$, $\beta$ and $\gamma$, given in (\ref{defBGA}), one has the identity
    \begin{equation}
        \alpha_{i}^{\phantom{i}k} + \gamma_{i}^{\phantom{i}k} = \delta_{i}^{\phantom{i}k}. 
    \end{equation}
    Using this relation, the expression (\ref{apptustustus}) is equivalently written in the form 
    \begin{equation}
        \label{APPTKKKK}
        [\overline{T}_i,\overline{K}_j] = i(\beta_{i}^{\phantom{i}l}\delta_{j}^{\phantom{j}m}-\alpha_{i}^{\phantom{i}l}\beta_{j}^{\phantom{j}m})\epsilon_{lm}^{\phantom{lm}n}(\underbrace{\alpha_{n}^{\phantom{n}k} + \gamma_{n}^{\phantom{n}k}}_{\delta_{n}^{\phantom{n}k}})K_k + i\alpha_{i}^{\phantom{i}l}\delta_{j}^{\phantom{j}m}\epsilon_{lm}^{\phantom{lm}n}(\underbrace{\alpha_{n}^{\phantom{n}k} + \gamma_{n}^{\phantom{n}k}}_{\delta_{n}^{\phantom{n}k}})J_k+i\beta_{i}^{\phantom{i}l}\beta_{j}^{\phantom{j}m}\epsilon_{lm}^{\phantom{lm}n}(\underbrace{\alpha_{n}^{\phantom{n}k} + \gamma_{n}^{\phantom{n}k}}_{\delta_{n}^{\phantom{n}k}})J_k.
    \end{equation}
    At this stage, we invoke the identities $\beta_{i}^{\phantom{i}l}\beta_{j}^{\phantom{j}m}\epsilon_{lm}^{\phantom{lm}n}\alpha_{n}^{\phantom{n}k} = 0$ and $\alpha_{i}^{\phantom{i}l}\delta_{j}^{\phantom{j}m}\epsilon_{lm}^{\phantom{lm}n}\alpha_{n}^{\phantom{n}k} = -(\beta_{i}^{\phantom{i}l}\delta_{j}^{\phantom{j}m}-\alpha_{i}^{\phantom{i}l}\beta_{j}^{\phantom{j}m})\epsilon_{lm}^{\phantom{lm}n}\beta_{n}^{\phantom{n}k}$. With these results, the commutator can be reorganized as
    \begin{eqnarray}
        \label{apphome}
        [\overline{T}_i,\overline{K}_j] &=& i(\beta_{i}^{\phantom{i}l}\delta_{j}^{\phantom{j}m}-\alpha_{i}^{\phantom{i}l}\beta_{j}^{\phantom{j}m})\epsilon_{lm}^{\phantom{lm}n}(\alpha_{n}^{\phantom{n}k}K_k-\beta_{n}^{\phantom{n}k}J_k) + i(\beta_{i}^{\phantom{i}l}\delta_{j}^{\phantom{j}m}-\alpha_{i}^{\phantom{i}l}\beta_{j}^{\phantom{j}m})\epsilon_{lm}^{\phantom{lm}n}\gamma_{n}^{\phantom{n}k}K_k +\nonumber\\
        &+& i(\alpha_{i}^{\phantom{i}l}\delta_{j}^{\phantom{j}m} + \beta_{i}^{\phantom{i}l}\beta_{j}^{\phantom{j}m})\epsilon_{lm}^{\phantom{lm}n}(\gamma_{n}^{\phantom{n}k}J_k).
    \end{eqnarray}
    The first and third terms in this expression can be immediately identified, respectively, with the definitions of $\overline{T}$ and $\overline{J}$. Consequently, the commutator takes the form
    \begin{eqnarray}
        \label{apphome2}
        [\overline{T}_i,\overline{K}_j] &=& i(\beta_{i}^{\phantom{i}l}\delta_{j}^{\phantom{j}m}-\alpha_{i}^{\phantom{i}l}\beta_{j}^{\phantom{j}m})\epsilon_{lm}^{\phantom{lm}k}\overline{T}_k + i(\beta_{i}^{\phantom{i}l}\delta_{j}^{\phantom{j}m}-\alpha_{i}^{\phantom{i}l}\beta_{j}^{\phantom{j}m})\epsilon_{lm}^{\phantom{lm}n}\gamma_{n}^{\phantom{n}k}K_k + i(\alpha_{i}^{\phantom{i}l}\delta_{j}^{\phantom{j}m} + \beta_{i}^{\phantom{i}l}\beta_{j}^{\phantom{j}m})\epsilon_{lm}^{\phantom{lm}k}\overline{J}_k.
    \end{eqnarray}
    Finally, to make the closure of the algebra explicit, we add a vanishing contribution, noting that
    \begin{equation}
        (\beta_{i}^{\phantom{i}l}\delta_{j}^{\phantom{j}m}-\alpha_{i}^{\phantom{i}l}\beta_{j}^{\phantom{j}m})\epsilon_{lm}^{\phantom{lm}n}\gamma_{n}^{\phantom{n}p}\beta_{p}^{\phantom{p}k} \equiv 0.
    \end{equation}
    This allows the second term to be rewritten as a contribution proportional to $\overline{K}$. Then, one obtains 
    \begin{eqnarray}
        \label{apphome3}
        [\overline{T}_i,\overline{K}_j] &=& i(\beta_{i}^{\phantom{i}l}\delta_{j}^{\phantom{j}m}-\alpha_{i}^{\phantom{i}l}\beta_{j}^{\phantom{j}m})\epsilon_{lm}^{\phantom{lm}k}\overline{T}_k + i(\beta_{i}^{\phantom{i}l}\delta_{j}^{\phantom{j}m}-\alpha_{i}^{\phantom{i}l}\beta_{j}^{\phantom{j}m})\epsilon_{lm}^{\phantom{lm}n}\gamma_{n}^{\phantom{n}k}\overline{K}_k+ \nonumber\\ &+& i(\alpha_{i}^{\phantom{i}l}\delta_{j}^{\phantom{j}m} + \beta_{i}^{\phantom{i}l}\beta_{j}^{\phantom{j}m})\epsilon_{lm}^{\phantom{lm}k}\overline{J}_k,
    \end{eqnarray}
    as we are meant to demonstrate. 

    Lastly, we consider $[\overline{T}_i,\overline{J}_j]$. As usual, from the definitions, one arrives at 
    \begin{equation}
        [\overline{T}_i,\overline{J}_j] = -i \alpha_i^{\phantom{i}l}\gamma_{j}^{\phantom{j}m}\epsilon_{lm}^{\phantom{lm}k}K_k + i\beta_i^{\phantom{i}l}\gamma_j^{\phantom{j}m}\epsilon_{lm}^{\phantom{lm}k}J_k.\label{s1}
    \end{equation}
    Taking into account the identities
    \begin{equation}
    \alpha_i^{\phantom{i}l}\gamma_{j}^{\phantom{j}m}\epsilon_{lm}^{\phantom{lm}k}=\gamma_{j}^{\phantom{j}m}\epsilon_{im}^{\phantom{im}n}\alpha_n^{\phantom{n}k}\qquad \text{and} \qquad \beta_i^{\phantom{i}l}\gamma_j^{\phantom{j}m}\epsilon_{lm}^{\phantom{lm}k} = \gamma_j^{\phantom{j}m}\epsilon_{im}^{\phantom{im}n}\beta_n^{\phantom{n}k},
    \end{equation}
    the relation (\ref{s1}) reduces to
    \begin{equation}
        [\overline{T}_i,\overline{J}_j] = -i\gamma_j^{\phantom{j}m}\epsilon_{im}^{\phantom{im}k}\overline{T}_k,
    \end{equation}
    by the identification of (\ref{3}). This completes the step-by-step derivation of Eqs. (\ref{AlgnKK}) -- (\ref{AlgnTJ}).

    \section{Detailed derivation of expression \ref{auidsauiddadasdas}}\label{App: B}
    We now cover in detail the steps that led us to expression (\ref{auidsauiddadasdas}). For convenience, we begin by defining the following notation:
    \begin{equation}
        \label{notaçãozinhamichu}
        \begin{cases}
            \overset{\circ}{\omega}_d{}^a{}_\mu = \omega^b{}_\mu \beta_b{}^c\epsilon_{cd}{}^a,\\
            \widetilde{\omega}_d{}^a{}_\mu = \frac{1}{2}\omega^{kl}{}_\mu\epsilon_{kl}{}^n\gamma_n{}^c\epsilon_{cd}{}^a.
        \end{cases}
    \end{equation}
    We start from the computation of $e^a{}_\sigma \widehat{R}^\sigma{}_{\rho\mu\nu}$, which can be represented as
    \begin{equation}
        e^a{}_\sigma\widehat{R}^\sigma{}_{\rho\mu\nu} = e^a{}_\sigma \partial_\mu \widehat{\Gamma}^\sigma{}_{\rho\nu} + e^a{}_\sigma \widehat{\Gamma}^\sigma{}_{\lambda \mu}\widehat{\Gamma}^\lambda{}_{\rho \nu} - (\mu\leftrightarrow \nu).
    \end{equation}
    Since we are in a torsionless scenario, one can reorganize the indices in the expression above as 
    \begin{equation}
        \label{curvintermed1}
        e^a{}_\sigma\widehat{R}^\sigma{}_{\rho\mu\nu} = e^a{}_\sigma \partial_\mu \widehat{\Gamma}^\sigma{}_{\nu\rho} + e^a{}_\sigma \widehat{\Gamma}^\sigma{}_{\lambda \mu}\widehat{\Gamma}^\lambda{}_{\nu\rho} - (\mu\leftrightarrow \nu).
    \end{equation}
    Then, making use of a convenient product rule
    \begin{eqnarray}
        e^a{}_\sigma \partial_\mu \widehat{\Gamma}^\sigma{}_{\nu\rho} &=& \partial_\mu (e^a{}_\sigma \widehat{\Gamma}^\sigma{}_{\nu\rho}) - (\partial_\mu e^a{}_\sigma) \widehat{\Gamma}^\sigma{}_{\nu\rho}\nonumber\\
        &=& \partial_\mu \left\{\partial_\nu e^a{}_\rho + \omega^a{}_\nu t_\rho - \overset{\circ}{\omega}_d{}^a{}_\nu e^d{}_\rho - \widetilde{\omega}_d{}^a{}_\nu e^d{}_\rho\right\} + \left\{\omega^a{}_\mu t_\sigma - \overset{\circ}{\omega}_d{}^a{}_\mu e^d{}_\sigma-\widetilde{\omega}_d{}^a{}_\mu e^d{}_\sigma   \right\}\widehat{\Gamma}^\sigma{}_{\nu\rho},
    \end{eqnarray}
    where we introduced the vielbein postulate (\ref{VBP2}) in the passage of the first to the second line, expression (\ref{curvintermed1}) reads 
    \begin{eqnarray}
        \label{curvintermed2}
        e^a{}_\sigma\widehat{R}^\sigma{}_{\rho\mu\nu} &=& \partial_\mu \left\{\partial_\nu e^a{}_\rho + \omega^a{}_\nu t_\rho - \overset{\circ}{\omega}_d{}^a{}_\nu e^d{}_\rho - \widetilde{\omega}_d{}^a{}_\nu e^d{}_\rho\right\} + \left\{\omega^a{}_\mu t_\sigma - \overset{\circ}{\omega}_d{}^a{}_\mu e^d{}_\sigma-\widetilde{\omega}_d{}^a{}_\mu e^d{}_\sigma   \right\}\widehat{\Gamma}^\sigma{}_{\nu\rho} + \nonumber\\
        &+& e^a{}_\sigma \widehat{\Gamma}^\sigma{}_{\lambda \mu}\widehat{\Gamma}^\lambda{}_{\rho \nu} - (\mu\leftrightarrow \nu).
    \end{eqnarray}
    Finally, resorting again to the vielbein postulate (\ref{VBP1}) and (\ref{VBP2}), Eq. (\ref{curvintermed2}) becomes
    \begin{eqnarray}
        \label{ticoticonofubá}
        e^a{}_\sigma\widehat{R}^\sigma{}_{\rho\mu\nu} &=& \left\{\partial_\mu \omega^a{}_\nu - \partial_\nu \omega^a{}_\mu +(\omega^d{}_\mu \overset{\circ}{\omega}_d{}^a{}_\nu - \omega^d{}_\nu\overset{\circ}{\omega}_d{}^a{}_\mu) + (\omega^d{}_\mu \widetilde{\omega}_d{}^a{}_\nu - \omega^d{}_\nu \widetilde{\omega}_d{}^a{}_\mu)\right\}t_\rho - \left\{\partial_\mu \widetilde{\omega}_d{}^a{}_\nu - \partial_\nu \widetilde{\omega}_d{}^a{}_\mu \right\}e^d{}_\rho -\nonumber\\
        &-& \mathcal{R}^a{}_{d\mu\nu}(\omega^a{}_\mu, \omega^{ab}{}_\mu) e^d{}_\rho, 
    \end{eqnarray}
    where
    \begin{equation}
        \label{wannafeel}
        \mathcal{R}^a{}_{d\mu\nu}(\omega^a{}_\mu, \omega^{ab}{}_\mu) = \partial_\mu \overset{\circ}{\omega}_d{}^{a}{}_{\nu} + \omega^a{}_\mu \omega_{d\nu} - \overset{\circ}{\omega}_c{}^a{}_\mu\overset{\circ}{\omega}_d{}^c{}_\nu - \overset{\circ}{\omega}_c{}^a{}_\mu\widetilde{\omega}_d{}^c{}_\nu- \widetilde{\omega}_c{}^a{}_\mu\overset{\circ}{\omega}_d{}^c{}_\nu-\widetilde{\omega}_c{}^a{}_\mu\widetilde{\omega}_d{}^c{}_\nu - (\mu\leftrightarrow \nu).
    \end{equation}
    Through a direct exercise, one verifies that expression (\ref{ticoticonofubá}) can be developed into
    \begin{eqnarray}
        \label{ticoticonofubá2}
        e^a{}_\sigma\widehat{R}^\sigma{}_{\rho\mu\nu} &=& R^a{}_{\mu\nu}(\widehat{K})t_\rho - \frac{1}{2}R^{kl}{}_{\mu\nu}(\widehat{M})\epsilon_{kl}{}^n\gamma_n{}^c\epsilon_{cd}{}^ae^d{}_\rho - \mathcal{R}^a{}_{d\mu\nu}(\omega^a{}_\mu, \omega^{ab}{}_\mu) e^d{}_\rho.
    \end{eqnarray}
    
    The treatment of the term containing $\mathcal{R}^a{}_{d\mu\nu}(\omega^a{}_\mu, \omega^{ab}{}_\mu)$, however, demands a deeper analysis. It is useful to first consider the following combination $R^c{}_{\mu\nu}(\widehat{K})\beta_c{}^n\epsilon_{nd}{}^a$, which by means of (\ref{curvK}), becomes
    \begin{eqnarray}
        \label{auxX}
        R^c{}_{\mu\nu}(\widehat{K})\beta_c{}^n\epsilon_{nd}{}^a &=& \partial_\mu \overset{\circ}{\omega}_d{}^a{}_\nu - \partial_\nu \overset{\circ}{\omega}_d{}^a{}_\mu + \omega^b{}_\mu\overset{\circ}{\omega}_b{}^c{}_\nu\beta_c{}^n\epsilon_{nd}{}^a - \omega^b{}_\nu\overset{\circ}{\omega}_b{}^c{}_\mu\beta_c{}^n\epsilon_{nd}{}^a + \omega^b{}_\mu \widetilde{\omega}_b{}^c{}_\nu \beta_c{}^n\epsilon_{nd}{}^a - \nonumber \\
        &-& \omega^b{}_\nu \widetilde{\omega}_b{}^c{}_\mu \beta_c{}^n\epsilon_{nd}{}^a.
    \end{eqnarray}
    Consequently, the introduction of the auxiliary expression (\ref{auxX}) into (\ref{wannafeel}) results in
    \begin{eqnarray}
        \mathcal{R}^a{}_{d\mu\nu}(\omega^a{}_\mu, \omega^{ab}{}_\mu) &=& R^c{}_{\mu\nu}(\widehat{K})\beta_c{}^n\epsilon_{nd}{}^a + (I)^a{}_{d\mu\nu} + (II)^a{}_{d\mu\nu} + (III)^a{}_{d\mu\nu},
    \end{eqnarray}
    where the following definitions are being used:
    \begin{eqnarray}
        \label{Iquasefim}
        (I)^a{}_{d\mu\nu} &\equiv& -\omega^b{}_\mu\overset{\circ}{\omega}_b{}^c{}_\nu\beta_c{}^n\epsilon_{nd}{}^a - \omega^a{}_\nu \omega_{d\mu} + \overset{\circ}{\omega}_c{}^a{}_\nu\overset{\circ}{\omega}_d{}^c{}_\mu - (\mu \leftrightarrow \nu)\nonumber\\
        &=& \omega^i{}_\mu \omega^j{}_\nu\left[-\beta_{j}{}^k\epsilon_{ki}{}^c\beta_{c}{}^n\epsilon_{nd}{}^a - \delta_j{}^a\eta_{id}+ \beta_j{}^n\epsilon_{nb}{}^a\beta_{i}{}^k\epsilon_{kd}{}^b - (i\leftrightarrow j) \right],\\
        \label{IIquasefim}
        (II)^a{}_{d\mu\nu} &\equiv& - \omega^b{}_\mu \widetilde{\omega}_b{}^c{}_\nu \beta_c{}^n\epsilon_{nd}{}^a - \overset{\circ}{\omega}_c{}^a{}_\mu\widetilde{\omega}_d{}^c{}_\nu + \widetilde{\omega}_c{}^a{}_\nu \overset{\circ}{\omega}_{d}{}^c{}_\mu\nonumber\\
        &=& \frac{1}{2}\omega^b{}_\mu \omega^{kl}{}_\nu\epsilon_{kl}{}^p\gamma_p{}^q\left(-\epsilon_{qb}{}^c\beta_c{}^n\epsilon_{nd}{}^a - \beta_b{}^m\epsilon_{mc}{}^a\epsilon_{qd}{}^c + \beta_b{}^m\epsilon_{md}{}^c\epsilon_{qc}{}^a \right) - (\mu\leftrightarrow \nu),\\
        \label{IIIquasefim}
        (III)^a{}_{d\mu\nu} &\equiv& - \widetilde{\omega}_c{}^a{}_\mu \widetilde{\omega}_d{}^c{}_\nu - (\mu\leftrightarrow \nu)\nonumber\\
        &=& \frac{1}{4}(\omega^{kl}{}_\mu\epsilon_{kl}{}^n\gamma_n{}^m) (\omega^{pq}{}_\nu\epsilon_{pq}{}^r\gamma_r{}^s)\left(-\epsilon_{mc}{}^a\epsilon_{sd}{}^c+\epsilon_{sc}{}^a \epsilon_{md}{}^c\right).   
    \end{eqnarray}
    Meanwhile, after analyzing equations (\ref{Iquasefim}) -- (\ref{IIIquasefim}) at the level of its components, one verifies the following identities:
    \begin{eqnarray}
        -\beta_{j}{}^k\epsilon_{ki}{}^c\beta_{c}{}^n\epsilon_{nd}{}^a - \delta_j{}^a\eta_{id}+ \beta_j{}^n\epsilon_{nb}{}^a\beta_{i}{}^k\epsilon_{kd}{}^b - (i\leftrightarrow j) &=& 0,\\
        \gamma_p{}^q\left(-\epsilon_{qb}{}^c\beta_c{}^n\epsilon_{nd}{}^a - \beta_b{}^m\epsilon_{mc}{}^a\epsilon_{qd}{}^c + \beta_b{}^m\epsilon_{md}{}^c\epsilon_{qc}{}^a \right) &=& 0 
    \end{eqnarray}
    and
    \begin{equation}
        \gamma_n{}^m\gamma_r{}^s\left(-\epsilon_{mc}{}^a\epsilon_{sd}{}^c+\epsilon_{sc}{}^a \epsilon_{md}{}^c\right) = 0.
    \end{equation}
    Therefore, expression (\ref{ticoticonofubá2}) simply reduces to
    \begin{equation}
        \label{penultimeiramaluca}
        e^a{}_\sigma\widehat{R}^\sigma{}_{\rho\mu\nu} = R^a{}_{\mu\nu}(\widehat{K})t_\rho - \frac{1}{2}R^{kl}{}_{\mu\nu}(\widehat{M})\epsilon_{kl}{}^n\gamma_n{}^c\epsilon_{cd}{}^ae^d{}_\rho - R^c{}_{\mu\nu}(\widehat{K})\beta_c{}^n\epsilon_{nd}{}^a e^d{}_\rho.
    \end{equation}
    As a last step, using $e^a{}_\sigma e_c{}^\sigma = \delta^a{}_c$, Eq. (\ref{penultimeiramaluca}) is recast as
    \begin{equation}
        e^a{}_\sigma\widehat{R}^\sigma{}_{\rho\mu\nu} = \left\{R^c{}_{\mu\nu}(\widehat{K})t_\rho e_c{}^\sigma  - \left(  R^b{}_{\mu\nu}(\widehat{K})\beta_b{}^n\epsilon_{nd}{}^c +\frac{1}{2}R^{kl}{}_{\mu\nu}(\widehat{M})\epsilon_{kl}{}^n\gamma_n{}^b\epsilon_{bd}{}^c\right)e^d{}_\rho e_c{}^\sigma \right\}e^a{}_\sigma,
    \end{equation}
    from which it can be recognized the desired result
    \begin{equation}
        \widehat{R}^\sigma{}_{\rho\mu\nu} = R^c{}_{\mu\nu}(\widehat{K})t_\rho e_c{}^\sigma  - \left( R^b{}_{\mu\nu}(\widehat{K})\beta_b{}^n\epsilon_{nd}{}^c  +\frac{1}{2}R^{kl}{}_{\mu\nu}(\widehat{M})\epsilon_{kl}{}^n\gamma_n{}^b\epsilon_{bd}{}^c\right)e^d{}_\rho e_c{}^\sigma.
    \end{equation}

\end{document}